\documentclass[twocolumn,twocolappendix,tighten]{aastex631} 
 \usepackage{mathptmx,courier}
  \usepackage[scaled=.92]{helvet}
  \usepackage{CJK}
  \DeclareMathAlphabet{\mathcal}{OMS}{cmsy}{m}{n}
\usepackage{natbib}
\usepackage{graphicx}
\usepackage[figuresright]{rotating}
\usepackage {amsmath,amssymb,eqnarray,tabularx}
\usepackage{enumerate}
\usepackage{enumitem}
\usepackage{makecell}
\usepackage{mathtools}
\hypersetup{bookmarksnumbered=true,bookmarksopen=true}
\usepackage[all]{hypcap}
\newcommand{\hi}{{\rm H}{\textsc i}}
\newcommand{\SN}{\ensuremath{ \text{S/N} } }

\newcommand{\degree}{\ensuremath{\text{\textdegree}}}
\ProvideDocumentEnvironment{CJK*}{m m}{}{}
\makeatletter
\AtBeginDocument{
  \newcommand*{\cjkname}[3][gbsn]{#2 (\begin{CJK*}{UTF8}{#1}#3\ignorespacesafterend\end{CJK*})}
  \@ifpackageloaded{CJK}{}{
    \@ifpackageloaded{xeCJK}{}{
      \iflatexml\else
        \renewcommand*{\cjkname}[3][]{#2}
      \fi
    }
  }
}
\makeatother

\def\jyb{\rm{Jy~beam^{-1} }}

\def\jybkms{\rm{Jy~beam^{-1}~km~s^{-1} }}

\def\mjyb{\rm{mJy~beam^{-1} }}

\def\kms{\rm{km~s^{-1} }}
\def\cmsq{\rm{ cm^{-2} } }
\def\NHI{N_{\rm HI}}
\def\lgNHI{\log(\NHI/\cmsq)}
\def\MHI{M_{\rm HI}}

\def\Msunpcsq{\rm{M_{\odot}} ~{\rm pc}^{-2}}

\defcitealias{Wang23}{W23} 
\defcitealias{Wang24}{W24}

\begin{document}


\title{ FEASTS and MHONGOOSE: HI Column Density Distribution at z$=$0 for $\NHI>10^{17.8}\, \cmsq$ }
\correspondingauthor{Jing Wang}
\email{jwang\_astro@pku.edu.cn}

\author[0000-0002-6593-8820]{\cjkname{Jing Wang}{王菁}}
\affiliation{ Kavli Institute for Astronomy and Astrophysics, Peking University, Beijing 100871, China}

\author{\cjkname{Xuchen Lin}{林旭辰}}
\affiliation{ Kavli Institute for Astronomy and Astrophysics, Peking University, Beijing 100871, China}

\author{\cjkname{Ze-Zhong Liang}{梁泽众}}
\affiliation{ Kavli Institute for Astronomy and Astrophysics, Peking University, Beijing 100871, China}

\author{W.J.G. de Blok}
\affiliation{Netherlands Institute for Radio Astronomy (ASTRON), Oude Hoogeveensedijk 4, 7991 PD Dwingeloo, the Netherlands}
\affiliation{Dept. of Astronomy, Univ. of Cape Town, Private Bag X3, Ronde-bosch 7701, South Africa}
\affiliation{Kapteyn Astronomical Institute, University of Groningen, PO Box 800, 9700 AV Groningen, The Netherlands}

\author{\cjkname{Hong Guo}{郭宏}}
\affiliation{Shanghai Astronomical Observatory, Chinese Academy of Sciences, Shanghai 200030, People's Republic of China}

\author{\cjkname{Zhijie Qu}{屈稚杰}}
\affiliation{Department of Astronomy and Astrophysics, The University of Chicago, 5640 S. Ellis Avenue, Chicago, IL 60637, USA}

\author{C{\'e}line P{\'e}roux}
\affiliation{European Southern Observatory, Karl-Schwarzschildstrasse 2, D-85748 Garching bei M{\"u}nchen, Germany}
\affiliation{Aix Marseille Universit{\'e}, CNRS, LAM (Laboratoire d'Astrophysique de Marseille) UMR 7326, F-13388 Marseille, France}

\author[0000-0001-7457-8487,gname=Kentaro,sname=Nagamine]{Kentaro Nagamine}
\affiliation{Theoretical Astrophysics, Department of Earth \& Space Science, Graduate School of Science, The University of Osaka, 1-1 Machikaneyama, Toyonaka, Osaka 560-0043, Japan}
\affiliation{Theoretical Joint Research, Forefront Research Center, Graduate School of Science, The University of Osaka, 1-1 Machikaneyama, Toyonaka, Osaka 560-0043, Japan}
\affiliation{Kavli IPMU (WPI), UTIAS, The University of Tokyo, Kashiwa, Chiba 277-8583, Japan}
\affiliation{Department of Physics \& Astronomy, University of Nevada, Las Vegas, 4505 S. Maryland Pkwy, Las Vegas, NV 89154-4002, USA}
\affiliation{Nevada Center for Astrophysics, University of Nevada, Las Vegas, 4505 S. Maryland Pkwy, Las Vegas, NV 89154-4002, USA}

\author{\cjkname{Luis C. Ho}{何子山}}
\affiliation{ Kavli Institute for Astronomy and Astrophysics, Peking University, Beijing 100871, China}

\author{\cjkname{Dong Yang}{杨冬}}
\affiliation{ Kavli Institute for Astronomy and Astrophysics, Peking University, Beijing 100871, China}

\author{Simon Weng}
\affiliation{Sydney Institute for Astronomy, School of Physics A28, University of Sydney, NSW 2006, Australia}
\affiliation{ARC Centre of Excellence for All-Sky Astrophysics in 3 Dimensions (ASTRO 3D), Australia}
\affiliation{ATNF, CSIRO Space and Astronomy,  PO Box 76, Epping, NSW 1710, Australia}

\author{Claudia del P. Lagos}
\affiliation{International Centre for Radio Astronomy Research, University of Western Australia, 35 Stirling Highway, Crawley, WA 6009, Australia}
\affiliation{ARC Centre of Excellence for All-Sky Astrophysics in 3 Dimensions (ASTRO 3D), Australia}
\affiliation{Cosmic Dawn Center (DAWN), Denmark}

\author{\cjkname{Xinkai Chen}{陈新凯}}
\affiliation{ Kavli Institute for Astronomy and Astrophysics, Peking University, Beijing 100871, China}

\author{George Heald}
\affiliation{CSIRO, Space and Astronomy, PO Box 1130, Bentley, WA 6102, Australia}

\author{J. Healy}
\affiliation{Netherlands Institute for Radio Astronomy (ASTRON), Oude Hoogeveensedijk 4, 7991 PD Dwingeloo, the Netherlands}
\affiliation{Jodrell Bank Centre for Astrophysics, School of Physics and Astronomy, University of Manchester, Oxford Road, Manchester M13 9PL, UK}
\affiliation{Department of Astronomy, University of Cape Town, Private Bag X3, Rondebosch 7701, South Africa}

\author{\cjkname{Qifeng Huang}{黄齐丰}}
\affiliation{ Kavli Institute for Astronomy and Astrophysics, Peking University, Beijing 100871, China}

\author{Peter Kamphuis}
\affiliation{ Ruhr University Bochum, Faculty of Physics and Astronomy, Astronomical Institute (AIRUB), 44780, Bochum, Germany }

\author{D. Kleiner}
\affiliation{Netherlands Institute for Radio Astronomy (ASTRON), Oude Hoogeveensedijk 4, 7991 PD, Dwingeloo, The Netherlands}

\author{\cjkname[bsmi]{Di Li}{李菂}} %
\affiliation{New Cornerstone Science Laboratory, Department of Astronomy, Tsinghua University, Beijing 100084, China}
\affiliation{National Astronomical Observatories, Chinese Academy of Sciences, 20A Datun Road, Chaoyang District, Beijing, People's Republic of China}

\author{\cjkname{Siqi Liu}{刘思琦}}
\affiliation{ Kavli Institute for Astronomy and Astrophysics, Peking University, Beijing 100871, China}
\affiliation{Department of the Internet of Things and Electronic Engineering, School of Computer and Communication Engineering, University of Science and Technology Beijing, Beijing 100083, China.}

\author{F. M. Maccagni}
\affiliation{INAF - Osservatorio Astronomico di Cagliari, Via della Scienza 5, 09047, Selargius, CA, Italy}

\author{Lister Staveley-Smith}
\affiliation{International Centre for Radio Astronomy Research, University of Western Australia, 35 Stirling Highway, Crawley, WA 6009, Australia}

\author{\cjkname{Zherong Su}{苏哲容}}
\affiliation{ Department of Astronomy, Peking University, Beijing 100871, China}

\author{Freeke van de Voort}
\affiliation{Cardiff Hub for for Astrophysics Research and Technology, School of Physics and Astronomy, Cardiff University, Queen's Buildings, The Parade, Cardiff CF24 3AA, UK}

\author{Fabian Walter}
\affiliation{Max Planck Institut f{\"u}r Astronomie, K{\"o}nigstuhl 17, D-69117 Heidelberg, Germany}
\affiliation{California Institute of Technology, Pasadena, CA 91125, USA; National Radio Astronomy Observatory, Pete V. Domenici Array Science Center, P.O. Box O, Socorro, NM 87801, USA}

\author{\cjkname{Fangxiong Zhong}{钟芳雄}}
\affiliation{ School of Physics, Peking University, Beijing 100871, China}

\author{\cjkname{Siwei Zou}{邹思蔚}}
\affiliation{Chinese Academy of Sciences South America Center for Astronomy, National Astronomical Observatories, CAS, Beijing 100101, China}
\affiliation{Departamento de Astronom{\'i}a, Universidad de Chile, Casilla 36-D, Santiago, Chile}

\begin{abstract}
We present the first $z=0$ $\hi$ column density distribution function, $f(\NHI)$, extending down to $\lgNHI=17.8$.
 This was derived from high-sensitivity 21-cm emission-line imaging at $\sim$1 kpc resolution.
 At high-column-densities (19.8$<\lgNHI<$21.3), our results align with earlier $z=0$ studies but benefit from 100 times greater sensitivity.
 Comparisons with $z\sim3$ quasar absorption-line studies reveal that $f(\NHI)$ at $z=0$ is systematically lower by 0.1-0.4 dex for $19.2<\lgNHI<21$.
 However, the distributions become comparable at $17.8<\lgNHI<19.2$, suggesting weak evolution in this regime.
Extrapolating the length incidence ($\mathrm{d}N/\mathrm{d}X$) for $\lgNHI>17.5$ implies a covering fraction ($f_\mathrm{cov}$) of $\sim0.7$ within 1-kpc-scale $\hi$-detected pixels at $z=0$.
Notably, for $17.8<\lgNHI<20$, impact parameters at a given $\NHI$ are significantly lower than previous $z\sim0$ absorption-line results and TNG50 simulation predictions.
This discrepancy indicates challenges in identifying galaxy counterparts for absorbers and in recovering low-column-density $\hi$ within cosmological simulations.
Finally, we derive a covering fraction of 0.006 for $\lgNHI>17.8$ gas within the virial radius around Milky-Way-like galaxies.
These findings provide new constraints on the baryonic flows and gaseous dynamics governing galaxy evolution.
\end{abstract}


\section{Introduction} 
\label{sec:introduction}
Galaxies need gas to form stars and grow.
The atomic hydrogen, $\hi$, is the major component of the interstellar medium (ISM) in star-forming galaxies, and serves as the reservoir of material for forming stars.
Only high-column density ($\lgNHI\gtrsim20.3$) $\hi$ in appropriate conditions can be efficiently converted to molecular gas and form stars \citep{Bigiel08, Bigiel10}.
This high-column-density $\hi$ is almost exclusively found in the ISM, and contributes $\gtrsim$90\% of the cosmic $\hi$ mass \citep[][Z05 hereafter]{Zwaan05}.
Yet, with a typical depletion time of 4 Gyr at $z=0$ \citep{Saintonge22}, and possibly even shorter at higher redshift \citep{Walter20}, this high-column-density $\hi$ was not in the ISM from the beginning, and does not stay there forever in the lifetime of a galaxy.
It is expected to be a dynamic reservoir, participating in the gas cycle of inflows, star formation and outflows \citep[e.g.,][]{Crain17}.
And it is the flow of gas traced by low-column density $\hi$ that has built and continuously replenishes this ISM reservoir \citep{Peroux20}.

Among the low-column densities, the range in $\NHI$ between $10^{17}$ and $10^{20}\,\cmsq$ is possibly the most relevant and interesting in the gas accretion scenario.
$\hi$ within this $\NHI$ range is commonly referred to as LLSs (Lyman Limit Systems).
QSO absorption at the Lyman limit has been the most important observational method over a wide range of redshifts.
Observational developments make $z=3$ the best redshift for LLS detections, though the situation is expected to improve with new instrumental efforts like the Blue-MUSE.
At $z=3$ both observations and simulations consistently suggest that LLSs are mostly in dark matter halos, and trace a large fraction of the inflowing gas, particularly the pristine one from the Inter-Galactic Medium (IGM) \citep{Fumagalli12, vandeVoort12}.
Gas traced by LLSs also includes static gas, outflowing gas, circulating gas, and tidally stripped gas, depicting a complex and dynamic picture of gas flows involving phase transitions \citep{vandeVoort12}.
A statistical measure of the LLS $\hi$ distribution therefore provides powerful constraints to the environment and driving physics of galaxy evolution.

The $\hi$ column density distribution metrics, including the distribution function, the incidence, the impact parameter, and the covering fraction, have helped develop macroscopic models and simulations of cosmological galaxy evolution \citep{Ribaudo11, Fumagalli13, Fumagalli14, vandeVoort19}, as well as recipes for sub-grid physical processes including feedbacks \citep{FaucherGiguere16, Crain17}, microscopic gas hydrodynamic processes in the circum-galactic medium (CGM) \citep{Ji20, Stern21} and radiative transfer involving the UV background (UVB) \citep{Nagamine10, Altay11, Rahmati13, Yajima12}.
Cosmological simulations reproduce the $\hi$ column density distribution function in the range $10^{17}$ to $10^{20}\,\cmsq$ at $z=3$ reasonably well \citep{vandeVoort12, Rahmati15}.

A major limitation with LLS observations is their restrictions to the QSO sight lines, making it impossible to contiguously map the gas distribution.
The association of LLSs to host galaxies is complex, due to galaxy clustering \citep{Qu23, Weng23}, faintness of low-mass galaxies that contain a large fraction of the $\hi$ \citep{Ribaudo11, Lagos14}, and the dynamic nature of the multiphase gas \citep{Chen20}.
Finally, LLS observations struggle to collect sufficient statistics for a clear interpretation at $z=0$ \citep{Peroux22}.
So far, no column density distribution function or length incidence for LLSs exist at $z=0$.
We cannot simply use the $\hi$ column density distributions observed at high-z (i.e. $z=0.75$, the lowest redshift for LLS incidence length deviation in \citealt{Fumagalli13}) to infer the one at $z=0$, as the hydrodynamic and radiative environment of galaxies have evolved significantly since cosmic noon \citep{Rahmati13, FaucherGiguere20}, in addition to the large-scale cosmic environment assembly.
Around 20 years ago, 21-cm emission line imaging was used to derive the $\hi$ column density distribution function and related metrics for Damped Lyman $\alpha$ (DLA) like systems at $z=0$ (Z05, \citealt{Braun12, RyanWeber03}).
Since then, the field has remained relatively quiet, and no new measurements at $z=0$ have been presented.

New radio telescopes recently made it possible to map the $\hi$ with around 1-kpc resolution down to the LLS column density regime.
It is found that, down to a column density limit of $10^{17.7}\,\cmsq$, the majority of detected $\hi$ is associated with dynamically cold galactic disks \citep{Wang24, Yang25}, exhibiting no evidence of connecting to cosmic filaments except for temporary tidal tails \citep{Lin25b,Marasco25}.
The characteristic radius for the 0.01$\,\Msunpcsq$ ($10^{18.1}\,\cmsq$) surface density correlates tightly with the $\hi$ mass but not with stellar or dark matter halo mass \citep{Wang25}.
These new observations imply that the statistical $\hi$ column density distribution at $z=0$ may differ from those at cosmic noon and earlier, where filamentary LLS structures may be prevalent \citep{vandeVoort12}.

The close association of $\gtrsim10^{17.7}\,\cmsq$ $\hi$ with galactic disks make it possible to count the cosmic column density distribution through weighting the galaxies according to the $\hi$ mass function (HIMF), a method firstly developed in \citet{RyanWeber03} for the DLA-like $\NHI$.
Contemporary developments in $\hi$ 21-cm statistical observations reveal that, the cosmic $\hi$ density, HIMF, and galactic $\hi$ mass fraction scaling relations have significantly evolved since $z=1$ \citep{Chowdhury20, Chowdhury22, Chowdhury22b}.
Even since $z=0.5$, the $\hi$ universe seems to have been actively evolving \citep{Bera22, Sinigaglia22, Pan23}.
It is therefore timely, feasible and necessary to revisit the cosmic $\hi$ column density distribution function and related metrics for the $z=0$ universe, with column density limits 100 times deeper than 20 years ago, and the physical focus switching from the DLA-like $\hi$ near molecular gas and star formation locations, to the LLS-like $\hi$ extending toward and interacting with the CGM.

We use data from FEASTS \citep[FAST Extended Atlas for Selected Targets Survey,][]{Wang23}, and MHONGOOSE \citep[MeerKAT $\hi$ Observations of Nearby Galactic Objects: Observing Southern Emitters,][]{deBlok24}.
These two surveys are the most sensitive 21-cm emission line imaging surveys at $z=0$ at present for statistical samples.
Section~\ref{sec:data} provides more details about the data, as well as supplementary interferometric data from existing surveys and archives.
These data are used to produce relatively uniform $\hi$ images with a column densities limit of $\sim10^{17.8}\,\cmsq$ at a resolution of 1 kpc in Section~\ref{sec:prepare_im}, which are then used to construct the $\hi$ column density distribution function ($f(\NHI)$) in Section~\ref{sec:contruct_fNHI}.
Section~\ref{sec:result} presents the main result, $f(\NHI)$ derived to a $\lgNHI$ limit of 17.8 at a resolution of 1 kpc for the $z=0$ Universe.
Section~\ref{sec:discuss} compares the $z=0$ $f(\NHI)$ to those at high redshifts $z\geq3$.
The length incidences and impact parameter probability distribution at a given $\lgNHI$ are also derived and compared to LLS results at $z=0$ and $z=3$ in section~\ref{sec:discuss}.
We assume a Kroupa IMF \citep{Kroupa01} in calculation of the stellar mass, and do not include helium in calculation of the $\hi$ mass.
Throughout the paper we adopt the $\Lambda$CDM cosmology with the $\Omega_{\Lambda}=0.7$, $\Omega_\mathrm{m} =0.3$, and $H_0=70 \mathrm{km} \mathrm{s}^{-1} \mathrm{Mpc}^{-1}$.

\section{Data and sample}
\label{sec:data}
\subsection{The FAST images of FEASTS}
\label{sec:data_feasts}
FEASTS obtained $\hi$ images for 55 $\hi$-richest Local-Volume galaxies that are above the Milky Way disk plane in FAST \citep[Five-hundred-meter Aperture Spherical radio Telescope]{Jiang19} observable sky using on-the-fly mode, during the period 2021.09-2024.06.

The selection of FEASTS targets start from a master catalog of $\hi$ fluxes, combining the $\hi$ catalogs of Local-Volume galaxies, including the ALFALFA extended catalog \citep{Hoffman19}, the catalog of \citet{Karachentsev13}, and the Hyperleda II catalog \citep{Paturel03}.
We select all the galaxies with $\hi$ flux $f_\mathrm{HI}>50$ Jy $\kms$, corresponding to a predicted $R_1>3.3'$ according to the $\hi$ size-mass relation (Wang et al. 2016), where $R_1$ is the radius where the azimuthally averaged $\hi$ surface density reaches $1\, \Msunpcsq$ ($\sim10^{20.1}\,\cmsq$).
Unperturbed $\hi$ disks should extend to $>2R_1$ at the targeted detection 3-$\sigma$ $\NHI$ limit of $10^{17.7}\cmsq$, based on the median $\hi$ radial profile of galaxies (Wang et al. 2025); perturbed $\hi$ disks can extend much further.
We further select galaxies with $0\degree<\delta<60\degree$  and Galactic latitude $|$b$|>20\degree$, and exclude M31 and M33 which are large galaxies in the Local Group.
These selections result in a parent sample of 118 galaxies.

A subset of 44 galaxies with $f_\mathrm{HI}>100$ Jy $\kms$ from the parent sample are taken as the high-priority sample and firstly observed.
These galaxies have a predicted radii of $R_1>9.3'$, so can be relatively well resolved by FAST but may have serious short-spacing problems in previous interferometric observations.
An additional random subset of 11 galaxies from the parent sample are also observed within the assigned time of the program.
They are slightly less extended than the high-priority sample, but still well resolved in the diffuse, low-column density $\hi$.
We find only one galaxy, NGC 4192, whose emission is blended with the Milky Way cirrus.
This cirrus was identified as large-scale ($\sim 0.5\degree$) structures in $\hi$ emission detected in continuous channels between -77 and 41 $\kms$.
NGC 4192 is therefore excluded, and the remaining observed sample of 54 galaxies are referred to as the observed FEASTS sample.

Each targeted galaxy in the observed FEASTS sample is mapped within a box region centered on the galaxy with a width L, where L is roughly the larger of $1\degree$ and 8 $R_1$, but the shape is further adjusted (enlarged in the horizontal or vertical direction) using knowledge from previous $\hi$ interferometric or single-dish observations.
Incidence of the radio frequency interference (RFI) contamination during the whole observation is minimal ($<6\%$).

The observing mode, data reduction and source finding procedures are described in details in \citet{Wang23, Wang24}.
We briefly summarize the key parameters and components here in the following.
The data reduction pipeline includes standard modules of RFI flagging, calibration, gridding, and further spectral baseline flattening.
The source finding procedure uses SoFiA \citep{Serra15,Westmeier21}, which identifies linked voxels above a threshold of 4-$\sigma$ after the cube is smoothed with a series of kernels that have a maximum size of 7 pixels (210$''$) in the x and y direction and 13 channels ($\sim20\kms$) in the z direction.

The products include data cubes, source masks, and moment maps.
The data cubes have a spatial resolution of $3.24'$ and a velocity resolution of 1.61 $\kms$.
An error map is derived for each moment-0 image, with the value in each pixel calculated as $\sigma n_\mathrm{ch}^{0.5} w_\mathrm{ch}$ in unit of $\jybkms$, where the $\sigma$ is the cube rms in $\jyb$ calculated as 3-$\sigma$ clipped $\sigma$ with voxels beyond the source masks, the $n_\mathrm{ch}$ the number of detected channels in the pixel line-of-sight, and $w_\mathrm{ch}$ the channel width in $\kms$.
The moment-0 images are converted to the $\hi$ column density images, and the depth $N_\mathrm{HI,lim}$ is estimated as the image contour level where the median $\SN=3$.
The median $N_\mathrm{HI,lim}$ is $10^{17.71}\,\cmsq$.

There are galaxies in interacting pairs or overlapping with small line-of-sight velocity differences.
Using the procedure described in \citet{Wang25} and \citet{Huang25}, we deblend the $\hi$ into different galaxies in 3D when possible, for the early-stage interacting systems and interlopers.
The 3D deblending procedure takes optical positions and radio or optical velocities when available from the NED (NASA/IPAC Extragalactic Database), and use them as initial guess for positions of galaxies.
It then utilizes the \textit{watershed} algorithm in the 3D to segment a data cube into individual galaxies.
As shown with mock tests in \citet{Huang25}, deblending in 3D is much more efficient and accurate in separating close galaxies than the traditional method in 2D.
There it was also shown that pairs that have projected distances larger than their average $R_1$ and radial velocity difference larger than twice their average rotational velocities can be deblended with a flux accuracy of $>90\%$.
As in Z05, isolated $\hi$ clouds are assigned to closest galaxies.

\subsection{The interferometric data to be combined with the FAST images of FEASTS}
The FEASTS team have developed a series of tools to combine the FAST and interferometric $\hi$ data.
The combined products effectively have the resolution of the interferometric data, and sensitivity close to that of the FAST data.
This section introduces the interferometric data to be used for combination with the FEASTS data in this study.

\subsubsection{HALOGAS}
As the deepest interferometric $\hi$ survey in the northern sky, HALOGAS \citep[Westerbork Hydrogen Accretion in LOcal GAlaxieS,][]{Heald11} contributes data for 12 galaxies in the FEASTS sample.
We take the high-resolution data cubes from the survey data release, which have a typical beam size of 20$''$, channel width of 4.12 $\kms$, and cube rms of 0.2 $\mjyb$.
We do not use the much deeper low-resolution data cubes, so as to achieve similar physical resolution as other interferometry data collected in this study (see below).

\subsubsection{THINGS}
THINGS \citep[The $\hi$ Nearby Galaxy Survey,][]{Walter08} has 13 galaxies in common with FEASTS, and three further in common with HALOGAS.
The galaxy NGC4449 has an $\hi$ structure significantly exceeding the primary beam of VLA (FWHM$\sim40'$), so we use alternative multi-pointing mosaics from the VLA archive (see below).
We therefore use THINGS data for 9 galaxies.
The survey data release shows that a natural-weighted data cube has typical beam size of 10 arcsec, channel width of 2.6 or 5.2 $\kms$, and cube rms of 0.4 $\jyb$.
We use CASA (Common Astronomy Software Applications) to multi-scale clean the calibrated and continuum-subtracted THINGS visibilities largely following the same procedure as described in \citet{Wang24}, but with an additional uv tapering of $10''$ and $20''$ for 5 and 4 galaxies, respectively.
This additional uv tapering helps the final data set achieve relatively uniform depth and physical resolution (see section~\ref{sec:sample_prop}).

\subsubsection{VIVA and LITTLE THINGS}
The VIVA \citep[VLA Imaging survey of Virgo galaxies in Atomic gas,][]{Chung09} and the LITTLE THINGS \citep[Local Irregulars That Trace Luminosity Extremes, The H{\sc i} Nearby Galaxy Survey,][]{Hunter12} each provide interferometric data for two galaxies.
The survey data releases show that a typical VIVA natural-weighted data cube has a beam size, channel width, and rms of 20$''$, 10.4 $\kms$, and 0.35 $\jyb$, and a typical LITTLE THINGS cube 10$''$, 1.3 or 2.6 $\kms$, and 0.5 $\jyb$.
We clean the calibrated and continuum-subtracted visibilities released by VIVA with a similar procedure as for THINGS, but without further uv tapering, as the resolution and sensitivity are already close to the target values.
We take the natural weighted data cubes from the LITTLE THINGS data release.
We do not clean the LITTLE THINGS data, as the multi-scale clean was already applied in the survey data release.
\citet{Hunter12} further conducted convergency tests to show that the residual fluxes are minimal after removing the convolved clean models from the dirty cubes.

\subsubsection{Other VLA archival data}
We obtain un-calibrated visibility data from the VLA archive for 16 galaxies.
Among this subsample, two galaxies (NGC 4449 and NGC 3628) have multi-pointing observations so that extended structures can be relatively well mapped in a mosaic.
These data are diverse in array configuration, observing time and RFI contaminating incidence, resulting in a relatively wide range of beam size and rms levels.
We reduce them with a python pipeline built with CASA scripts following standard VLA 21-cm emission line data reduction procedures (Z. Liang et al. in prep).
The clean procedure is similar to that described above for the surveys conducted at VLA.

\subsubsection{The final FEASTS dataset}
The interferometric datasets described above make the initial FAST+interferometry sample, which contains 45 galaxies. %
Figure~\ref{fig:intprop} in Appendix summarizes the resolution and $\NHI$ sensitivity for all the interferometric data available from existing surveys or reduced by this team from the VLA archive.
The column density sensitivity $N_\mathrm{HI,lim}$ is estimated as the 3-$\sigma$ limit assuming a line width of 20 $\kms$.
The interferometric data for the seven galaxies that have resolution better than 0.8 kpc are degraded to 0.8 kpc resolution.
The six galaxies with interferometric data resolution poorer than 2.4 kpc (an arbitrary criteria) are excluded from the initial FAST+interferometry sample.
So the final FAST+interferometry sample (the {\it FEASTS sample} for short hereafter) has 40 galaxies.
The $N_\mathrm{HI,lim}$ is between $10^{18.6}$ and $10^{19.8}\,\cmsq$.
The $N_\mathrm{HI,lim}$ and resolution for these 40 galaxies are further summarized in Table~\ref{tab:intdata} in Appendix.

We use SoFiA to conduct source finding from these interferometric data cubes.
We use the smooth-and-clip finder, and a series of smoothing kernels with maximum sizes equal to the beam major axis in the x and y directions, and the higher of $\sim20\,\kms$ and 2 channels in the z direction.
The smoothing kernels in the x and y direction are set to be relatively conservative, so to minimize the contamination of noise in the final moment maps.

\subsection{MHONGOOSE} %
MHONGOOSE targets 30 nearby star-forming galaxies in the southern sky with MeerKAT \citep{deBlok24}.
The targeted galaxies have distances between 3 and 23 Mpc, and have been selected to have a relative uniform $\log \MHI$ distribution.
The survey produces data cubes and moment maps for galaxies, at a series of different resolutions and column density sensitivities, ranging from 10$^{17.8}\,\cmsq$ at 90$''$ resolution and $10^{19.7}\,\cmsq$ at 7$''$ resolution, for a line width of $20\,\kms$.
At the distances of these galaxies, the resolution of 20$''$ (the r10\_t00 data) correspond to a physical resolution similar to that of the FEASTS+interferometry sample.
The best column density sensitivity at 90$''$ (the r10\_t90 data) is very close to that of FEASTS.

\begin{figure*}
  \centering
  \includegraphics{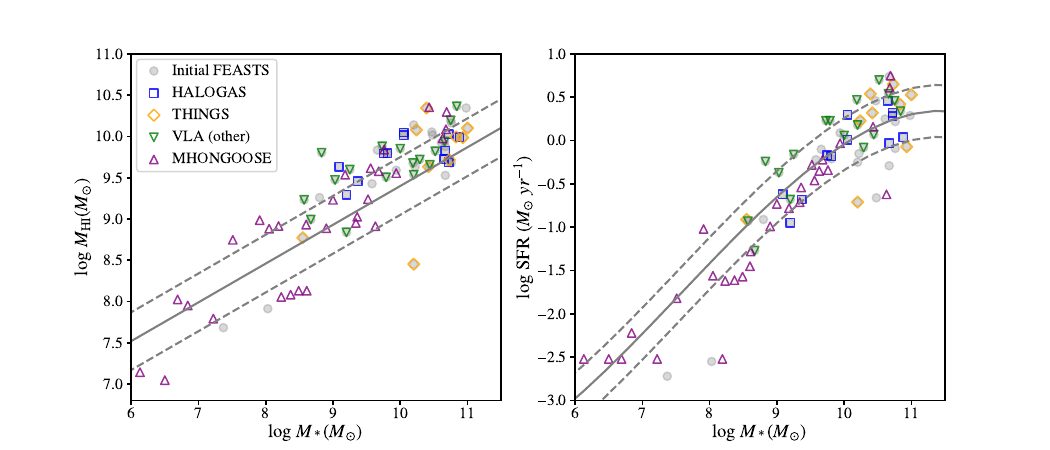}
  \caption{ The location of the final sample in space of $\MHI$ or SFR versus  $M_*$. The solid and dashed lines in the left panel are for the star forming main sequence (SFMS) and scatter \citep{Saintonge16}, and those in the right are for the median relation and scatter of galaxies on the SFMS \citep{Catinella18}. The squares, diamonds, and downward triangles plot the FEASTS sample for which the interferometric data are taken from the HALOGAS, the THINGS and other VLA surveys or archive, respectively. The observed FEASTS sample of 55 galaxies are also plotted as grey dots for reference. The upward triangles plot the MHONGOOSE sample; one MHONGOOSE galaxy (J1321-31) has $M_*=10^{4.7}\,M_{\odot}$ and $\MHI=10^{7.6}\,M_{\odot}$, and is thus beyond the plotted range. }
  \label{fig:scaling}
\end{figure*}

\subsection{The final main sample} 
\label{sec:sample_prop}
The final main sample is the FEASTS sample of 40 galaxies plus the MHONGOOSE released sample of 30 galaxies, in total 70 galaxies.
For the whole main sample, the galaxy distances are taken from the z0mgs ($z=0$ Multiwavelength Galaxy Synthesis, \citealt{Leroy19}).
The star formation rate (SFR) and stellar mass ($M_*$) are taken from the z0mgs catalog and \citet{deBlok24} for the FEASTS and MHONGOOSE sample, respectively.

The SFR, $M_*$ and $\hi$ mass ($M_\mathrm{HI}$) of the MHONGOOSE and FEASTS samples are displayed in Figure~\ref{fig:scaling}, with different symbols denoting the different samples of galaxies.
The MHONGOOSE and FEASTS samples are dominated by dwarf and Milky-Way like galaxies respectively, but together they cover a relatively wide range of $M_*$ and $M_\mathrm{HI}$.
MHONGOOSE is more evenly distributed in $\log M_\mathrm{HI}$ range, but has relatively sparse sampling, particularly at the low and high-$M_\mathrm{HI}$ ends.
FEASTS is biased toward the high-$M_\mathrm{HI}$ end, but helps fill $M_\mathrm{HI}$ bins throughout the whole $M_\mathrm{HI}$ range between $10^{7.5}$ and $10^{10.5}\,M_{\odot}$.
These features can also be seen from Figure~\ref{fig:full1kpchist}-e.
Both samples are biased toward highly star-forming and $\hi$-rich galaxies, a bias to keep in mind.

\section{{Preparing the HI images}}
\label{sec:prepare_im}
\subsection{Combining FAST and interferometric data into full-HI images} 
\label{sec:combine_fst}
We combine FAST and interferometric images into a single image, to achieve both high sensitivity and high resolution.
All interferometric images used for combination have been corrected for primary beam attenuation.
In each image, the interferometric data contribute the high-column-density and small-scale signals (referred to as the dense $\hi$), while the initial FAST data only contribute the remaining diffuse $\hi$ that has intrinsically large scales.
The large $\hi$ angular sizes for the diffuse $\hi$ are guaranteed by sample selection, and further supported by the characteristic radius $r_{001}$ at a surface density of $0.01\,\Msunpcsq$ ($\sim10^{18.1}\,\cmsq$), being greater than $9'$, and 2.8-8 times the beam major axis (FWHM of PSF) in the FEASTS sample \citep{Wang25}.
When the diffuse $\hi$ has a large angular size that is well resolved by the FAST beam, the combined images are effectively the full-resolution $\hi$ images.


\subsubsection{Preparations for combination}
As the nature of the combination method is feathering, imaging artifact in the interferometric imaging data cannot be fixed and propagates to the creation of full-resolution column density map.
To minimize this effect, all the interferometric cubes used in this study are visually inspected to ensure that there are no significant imaging artifacts.

As described in \citet{Wang24} and \citet{Wang24b}, the moment-0 images are used for the combination, to suppress systematic errors due to noise and unresolved structures in individual channel maps.
From the interferometry side, moment-0 images derived from masked standard cubes (convolved clean models plus dirty residuals) are used when the data is from the HALOGAS or LITTLE THINGS, as these two datasets have high SNRs and relatively zero-level residuals.
Otherwise, moment-0 images from the convolved models (without residuals being added) are used for combination, to avoid errors due to difference in clean and dirty beam areas \citep{Wang24}.

We cross-calibrate the flux levels between the FAST and interferometric images of each galaxy, using the procedure described in \citet{Wang24}.
In the flux cross-calibration procedure, the two types of data are compared in uv overlapping regions in the Fourier space, to derive the best-fit linear relation between them.
The average factor for interferometric data to scale in order to match the FAST flux calibration is 1.04$\pm$0.06, consistent with typical radio flux calibration errors of 10\%.

We also cross-calibrate the WCS systems of the two types of images, as described in \citet{Wang24}.
The WCS correction to the FAST image is determined as shifts in the WCS parameters CDELT1 and CDELT2 that lead to the lowest standard deviation in the difference image between the FAST and PSF matched interferometric images.
The average WCS shift is around 8$"$ in both $x$ and $y$ directions, consistent with the typical pointing error of 10$''$ for FAST \citep{Jiang19}.

In the next two sections, we use two different procedures, both of which can be classified as the linear method \citep{Stanimirovic02}, to combine the FEASTS and interferometric images.

\subsubsection{The traditional combination procedure}
\label{sec:combine_feasts_traditional}
The traditional procedure largely follows the workflow of \citet{Stanimirovic02} or IMMERGE of MIRIAD, and has been optimized for the FEASTS data to account for the average FAST beam shape in \citet{Wang24b}.
The procedure transforms both images to the Fourier space, combines the two types of data with a weighting function, and then inverse-transform the combined data back to the image domain.
The weighting function is solved by requiring the (FAST average beam + interferometric synthesized beam) combined beam to be equivalent to the interferometric synthesized beam.
The inverse-transformed image is the final combined image.

Mathematically, what the procedure achieves is equivalent to convolving an interferometric image to have the PSF of the corresponding FEASTS image, obtaining the difference between the convolved interferometric image and FEASTS image (i.e., the diffuse $\hi$), and adding this difference image back to the original interferometric image.

A possible problem associated with the first method is that the average beam is an imperfect representative of the FAST beams.
The FAST imaging, like that of many other single-dish radio telescope utilizing multi-beam receivers, suffer from variation of the multiple beams \citep{Jiang19, Chen25}, and non-Gaussian shape of the average beam due to light obscuration and refraction \citep{Wang23}.
These problems are most strong with the side-lobes of the FAST beams at the level of $\sim5\%$ \citep{Jiang19, Wang23,Chen25}, so they add systematic uncertainties to the low column densities, complicating interpretation of the faint $\hi$ structures.
The problems are much mitigated after the traditional combination procedure, because the sidelobe effects are shifted to much lower column densities in a diffuse $\hi$ map than in a full $\hi$ map.
But they cannot be ignored if the diffuse $\hi$ has a peak column density $\sim$20 times higher than the image column density limit ($>10^{19}\,\cmsq$), which is often the case when combining with typical interferometric data \citep{Wang24}.

\subsubsection{The new combination procedure}
\label{sec:combine_feasts_new}

We design procedures to tackle the problems related to the imperfect average beam of FAST images.
The goal is that, in the final full $\hi$ image, the diffuse $\hi$ is resolved with
a uniform and clean (Gaussian) PSF with FWHM close to that of the FAST average beam obtained in \citet{Wang23}.
The procedure has two crucial parts: accurate derivation of the diffuse $\hi$, and beam correction for the diffuse $\hi$.

In our previous work, the diffuse $\hi$ image is obtained by simply convolving the interferometric image with the FAST average beam, convolving the FAST image with the interferometric synthesized beam, and then subtracting the former from the latter \citep{Wang24}.
Here, the interferometric image is effectively convolved in a different way, through the FAST observing simulator described in \citet{Yang25}.
The simulator mimics the tracking and response of each beam in the 19-beam receiver of FAST during real observation of the target, but uses the interferometric image as the mock sky.
We then use the same pipeline that produces FAST $\hi$ images to grid the simulated observation into an $\hi$ image.
The convolved FAST image minus the FAST-simulator processed interferometric image produces the diffuse $\hi$ image here.

High-quality FAST beam images are essential to perform the overall procedure described above.
The beam images were taken during the FAST commissioning \citep{Jiang19}, and have a dynamic range of 3.3 dex within a radius of $\sim9'$ \citep{Wang23}, sufficient for sampling the $\NHI$ range from the FEASTS limit $10^{17.8}$ up to the ``saturating'' column density $10^{21.1}\,\cmsq$, with the angular size $r_{001}>9'$.
For worst cases of edge-on ($i>90\deg$) galaxies that have high $\NHI$ in the center and small angular sizes along the minor axis, \citet{Yang25} showed that the $\NHI$ minor-axis profiles measured from FEASTS images show clear excess over the average beam profile and its extrapolation, ensuring that no low-$\NHI$ emission erroneously being introduced in (or taken away from) the diffuse $\hi$ images.
Thus, on the whole, the FAST beam images plus the FEASTS $\hi$ images, in combination with the FAST observing simulator, allow relatively accurate extraction of the diffuse $\hi$ images.

Then, we only need to correct for the beam variation and non-Gaussian problems in the diffuse $\hi$ image.
The corrected diffuse $\hi$ image plus the original interferometric image will produce the {\it full $\hi$ image}.
We do not correct for the beam problems directly on the FAST images at an earlier step, because the systematic bias (and the uncertainty of correcting for them) due to an imperfect beam should be proportional to the surface brightness of the signals.
Therefore, working on the diffuse $\hi$ images suppresses the uncertainties that would be introduced by image beam corrections.

In the following, we introduce the procedures that unify and clean the beam in the diffuse $\hi$ image.

\begin{itemize}
\item {\bf Beam unification.}
As demonstrated with mock tests in Appendix~\ref{app:mock}, the beam variation problem is coupled with and amplified by the gridding process in image reduction, and shows up as sharp grid-like features in the final image.
Therefore, it is possible to conduct ``re-observation'' to further amplify and separate these high-order features.

We use the original diffuse $\hi$ image as the mock sky for the FAST observing simulator, and produce the re-observed image.
We subtract the re-observed image from the original diffuse $\hi$ image, and the resulting difference image serves as the correction image.
Then, the correction image is subtracted from the original diffuse $\hi$ image.
This corrected image has a beam similar to the average beam of FAST \citep{Wang23}, and is taken as the beam unified image.
The success of beam unification through this method is supported by mock tests in Appendix~\ref{app:mock}.

\item {\bf Beam clean.}
The original average beam deviates in shape from a 2D Gaussian function.
Making the beam Gaussian-like is equivalent to the clean procedure commonly used in interferometric data reduction.
We use the maximum likelihood method \citep{Richardson72, Lucy74} to conduct the deconvolution, through a python script converted from its IDL version in IDL Astronomy User's Library.
We also tried the maximum entropy method \citep{Agmon79} with a python script converted from the same IDL library, but the results are less satisfactory and this method is not adopted in the end.
Like the typical ``clean'' procedure, the residual image (input image minus the PSF-convolved model image) is derived in each iteration to assess the goodness of model.

We set 2-$\sigma$ as threshold of convergency for peak residual fluxes within the detecting area of the input image, and 5000 as the maximum number of iterations.
Convergence is typically reached within 5000 iterations.
We visually inspect the final residual images to ensure the clean quality.
The effectiveness of this beam clean procedure is also supported by the mock tests in Appendix~\ref{app:mock}.
\end{itemize}

The mock tests in Appendix~\ref{app:mock} also show that the diffuse $\hi$ surface density is reproduced with a relative uncertainty of $\lesssim$0.04 dex.

\subsubsection{The full-HI images of FEASTS}
The full-$\hi$ images combined with the new procedure are used for main-stream analysis in this study.
The main difference from the full-$\hi$ image produced with the traditional procedure is that the diffuse $\hi$ does not have significant beam variations throughout the image, or significant beam side lobes with respect to the Gaussian beam.

\begin{figure*}
  \centering
  \includegraphics{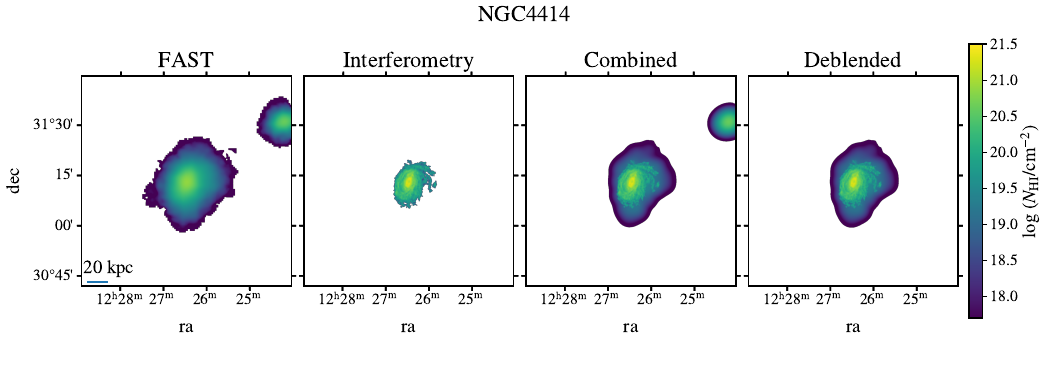}
    \includegraphics{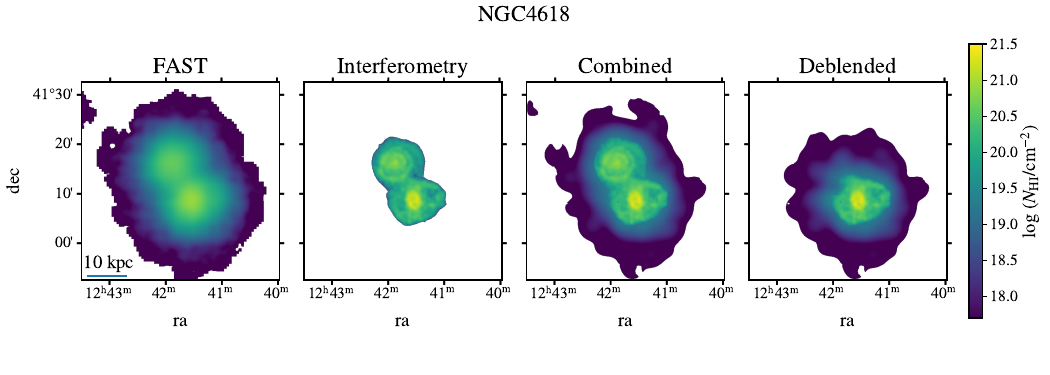}
  \caption{ An atlas of column density maps for galaxies in the FEASTS sample. For each galaxy, from left to right, we display the column density maps from the FAST data, the interferometric data, the combined full-$\hi$ image, and the combined and deblended full-$\hi$ image, respectively. }
  \label{fig:atlas_feasts}
\end{figure*}

An example atlas of these full-$\hi$ images are displayed in Figure~\ref{fig:atlas_feasts}, in comparison to the interferometric and FAST-only images.
The complete atlas can be found as supplementary material on-line.
But we also compared to results obtained with the traditional combination procedure, as an estimate for the 19-beam systematic effects when the new procedure is not applied.

\subsection{Constructing MHONGOOSE full-HI images}
\label{sec:combine_mg}
We combine the MHONGOOSE moment-0 images of different resolutions, in a way similar to the FEASTS+interferometric method.
All the moment-0 images used have been corrected for primary beam attenuation, and are thresholded by requiring $\SN>3$ before the following operation.
The error images used to derive the SNR are obtained with the same procedure as in section~\ref{sec:data_feasts}.

The r10\_t00 data has a similar angular and physical resolution as the interferometry dataset of FEASTS, and is taken as the highest-resolution signal in the combination.
Starting from r10\_t00 data, there are three types of data with successively lower resolution and higher sensitivity: r15\_t00, r05\_t60, and r10\_t90.
For each pair of successively two types, we smooth the higher-resolution image to the resolution of the lower-resolution one, and subtract it from the latter.
This difference image presents more diffuse signal uniquely detected in the lower-resolution image in the pair.
In total, 3 difference images are derived for the 3 pairs of successive data types.
They are all added to the highest-resolution image, which produces the full-$\hi$ image.
The full-$\hi$ images have the resolution of the r10\_t00 images ($\sim10"$), and the $N_\mathrm{HI,lim}$ of the r10\_t90 images ($\sim10^{18}\,\cmsq$)

\begin{figure}
  \centering
  \includegraphics{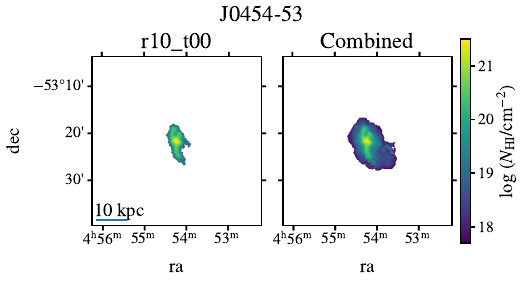}
  \includegraphics{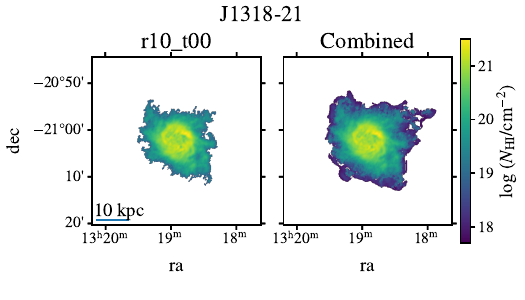}
  \caption{ An atlas of column density maps for galaxies in the MHONGOOSE sample. For each galaxy, we display the column density maps from the r10\_t00 data and the combined full-$\hi$ image.
 }
  \label{fig:atlas_mg}
\end{figure}

This procedure in principle is similar to the full-depth map construction introduced in \citet{deBlok24}.
In \citet{deBlok24}, the full-depth maps are produced mainly to display in one plot the contour levels uniquely revealed in images of different sensitivities and resolutions.
The main difference here from their method is that the contribution to low-$\NHI$ regions of scattered light from high-$\NHI$ regions is better accounted for.
This approach may be sensitive to variations in the synthesised beam shapes which will range from Gaussian at high SNR to non-Gaussian at low SNR due to the deconvolution threshold.
Because MeerKAT has a relatively flat noise level as a function of resolution, and MHONGOOSE has relative conservative threshold in clean and source finding processes \citep{deBlok24}, we assume this error to be relatively small.
We present an example atlas of these MHONGOOSE full-$\hi$ images in Figure~\ref{fig:atlas_mg}, in comparison to the original r10\_t00 images.

\section{Constructing the $f(\NHI)$ to $\NHI\sim10^{17.8}\,\cmsq$ }
\label{sec:contruct_fNHI}
In the following, we firstly summarize properties of the main sample (FEASTS+MHONGOOSE) and image data, to characterize the resolution and $\NHI$ range of $f(\NHI)$ to be constructed, and understand possible limitations.
Then we explain the construction procedure for $f(\NHI)$ (section~\ref{sec:construct_fNHI_procedure}).
Finally, we discuss systematic uncertainties (section~\ref{sec:construct_fNHI_error}), and possible systematic dependence of $f(\NHI)$ on sample and resolution, in light of comparison between studies (section~\ref{sec:result_systematics}).

\subsection{Data quality in the context of $f(\NHI)$ construction}
\label{sec:construct_fNHI_data}
\begin{figure*}
  \centering
  \includegraphics{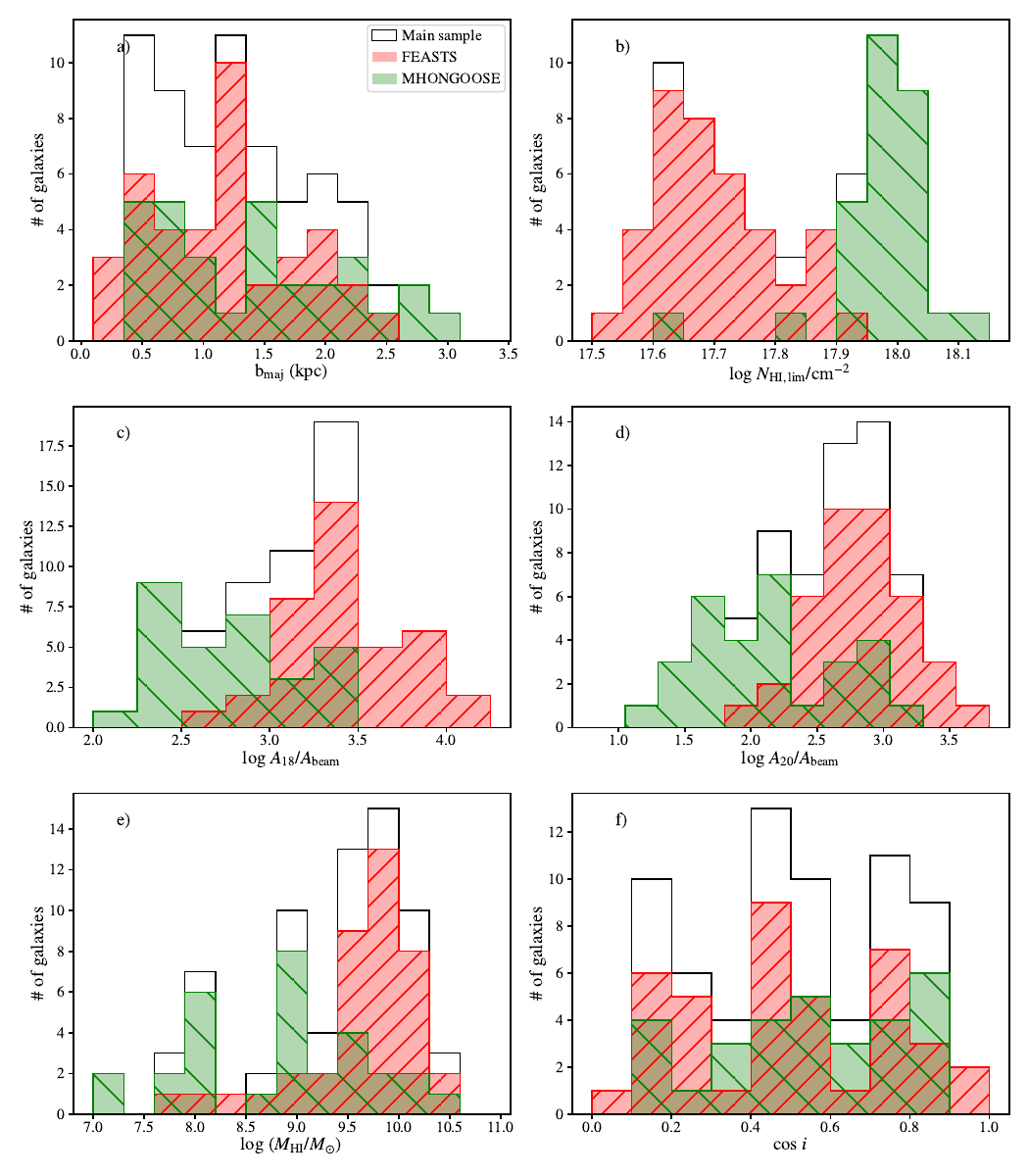}
  \caption{ Properties of the main sample that are most relevant for $f(\NHI)$ constructions. Panels a to f show the number distributions of galaxies in properties including image physical resolution (beam major axis in unit of kpc), image $\hi$ column density limit, the number of beam elements for $\lgNHI>18$, the number of beam elements for $\lgNHI>20$, the $\hi$ mass, and the galactic axis ratio. The distributions of the main sample (MS) are plotted as open histograms with black outlines. Distributions of the FEASTS sample are plotted in red histograms with upward diagonal stripes, and those of the MHONGOOSE sample in green histograms with downward diagonal stripes.
 }
  \label{fig:full1kpchist}
\end{figure*}

Figure~\ref{fig:full1kpchist}-a shows that, the median spatial resolution of the main-sample data is 1.2 kpc, and Figure~\ref{fig:full1kpchist}-b shows that the 50 and 90-percentile of column density limit $N_\mathrm{HI,lim}$ is $10^{17.8}$ and $10^{18.0}\,\cmsq$.
These statistics suggest that the $f(\NHI)$ constructed with the main sample is 50\%-complete down to $10^{17.8}\,\cmsq$ at $\sim$1-kpc resolution.

In order to accurately construct $f(\NHI)$, the area counts of each $\NHI$ should be statistically large, the images should spatially resolve the $\hi$ structures, the $\MHI$ coverage of sample should be sufficiently wide and complete corresponding to the bulk of cosmic $\hi$ density according to the HIMF, and the inclination distribution should be relatively random (cos$\,i$ following a uniform distribution) so to minimize biases caused by projection.
With these considerations, Figure~\ref{fig:full1kpchist}-c to d evaluate the feasibility of using the main sample images in constructing the $f(\NHI)$.

Figures~\ref{fig:full1kpchist}-c and d show the number of $\NHI$ counts contributed by each galaxy with its available data above a column density level of $10^{18}\,\cmsq$ and $10^{20}\,\cmsq$ respectively.
We have sufficient statistics putting the whole galaxy sample together for both $\lgNHI>18$ and $\lgNHI>20$.
The $\NHI>10^{18}\,\cmsq$ areas are well resolved for individual galaxies throughout the main sample, with 10 and 90 percentiles of 245 and 1695 beams (Figures~\ref{fig:full1kpchist}-c).
The $\NHI>10^{20}\,\cmsq$ areas are also reasonably resolved in individual galaxies, with a minimum of 10 beams (Figures~\ref{fig:full1kpchist}-d).
On the other hand, the areas of $\NHI>10^{21.3}\,\cmsq$ are relatively poorly resolved, being sampled by $<$30 beams in 23 galaxies (not plotted).

Figure~\ref{fig:full1kpchist}-e shows that the $\MHI$ ranges from $10^{7}$ to $10^{10.7}\,M_{\odot}$, and we have at least one galaxy in each $\MHI$ bin of 0.3 dex in the range $10^{7.5}$ to $10^{10.7}\,M_{\odot}$.
The cosmic $\hi$ density contributed by the $10^{7.5}$ to $10^{10.7}\,M_{\odot}$ range can be estimated as $\sum_{i} \psi(M_i) \omega(M_i) M_i$ \footnote{For the $i$-th $\MHI$ bin in the distribution, $M_i$, the $\psi_i$ is the HIMF value at $M_i$, and the $\omega(M_i)$ is the reciprocal of galaxy number in the $\log \MHI$ bin per the bin length of 0.3 dex (see section~\ref{sec:construct_fNHI_procedure})}.
The result recovers 100.0\% of the whole cosmic $\hi$ density $\Omega_\mathrm{HI}$ reported in \citet{Guo23}, suggesting sufficient coverage in $\MHI$.
The statistics seems overabundant at the high-$\MHI$ end, but can be useful as the larger $\hi$ disks can be more perturbed and show larger variance between each other in $\NHI$ histograms.

Figure~\ref{fig:full1kpchist}-f shows that the cos$\,i$ distribution is roughly flat.
It is not perfectly uniform, but not strongly tilted toward 0 or unity, throughout the sample.
Ensuring this rough flatness is important because more inclined galaxies tend to contribute more high-$\NHI$ areas than low-$\NHI$ ones due to projection.

To summarize, the main sample is sufficient for constructing a first $f(\NHI)$ down to the $10^{17.8}\cmsq$ level at a resolution of $\sim$1 kpc.
The relatively small main sample combined with galaxy diversity may cause systematic uncertainties, which may be worsened by the limited number of $\lgNHI>21.3$ elements and limited number of high-$i$ galaxies.
We will thus quantify these uncertainties, together with uncertainties from distance errors and other factors, in section~\ref{sec:construct_fNHI_error}.

\subsection{The $f(\NHI)$ construction procedure} 
\label{sec:construct_fNHI_procedure}
We largely follow the procedure of Z05 to construct the column density distribution $f(\NHI)$ with the main sample.

The moment-0 images are converted to column density images with an optically thin assumption.
The angular area of a column density image in each $\lgNHI$ bin is counted, and the $\lgNHI$ bins are set to range from 17.5 to 22.5 with a step of 0.1.
The area ($A_i$) measured from each galaxy is assigned a weight proportional to the HIMF $\psi_i$ evaluated at its $\hi$ mass ($M_i$), and inversely proportional to the number of galaxies in a $\log \MHI$ bin per the bin length (the $\omega(M_i)$).
The $\log (\MHI/M_{\odot})$ bins are set to range from 7 to 10.7 with a size of 0.3.
The $f(\NHI)$ is calculated as
\begin{equation}
f(\NHI)=\frac{c}{H_0} \frac{\sum_{i} \psi(M_i) \omega(M_i) A_i(\log \NHI) }{\NHI \ln 10 \Delta \log \NHI }
\end{equation}
As in Z05, we calculate errors propagated from the HIMF parameters, and contributed by Poisson errors of area counts in each $\NHI$ bin.

The major differences from the Z05 procedure are that
\begin{enumerate}
\item The reliable $\lgNHI$ bins start from 17.8 instead of 19.8 (section~\ref{sec:construct_fNHI_data}).
\item The HIMF has been updated to the latest one constructed with the ALFALFA data \citep{Guo23}.
Because the new HIMF has much lower parameter errors than the one of \citet{Zwaan03} used in Z05, it propagates to much lower errors in the achieved $f(\NHI)$.
\item The error bars additionally take into account the distance uncertainties and stochastic sampling of diverse galaxies.
Consideration of these additional uncertainties is motivated by the sample properties discussed in section~\ref{sec:construct_fNHI_data}.
As will be explained in section~\ref{sec:construct_fNHI_error}, these uncertainties contribute to $\gtrsim80\%$ of the $f(\NHI)$ variance throughout the $\NHI$ range.
\end{enumerate}

\subsection{Systematic uncertainties}
\label{sec:construct_fNHI_error}

\subsubsection{Systematic uncertainties due to distance errors and random sampling of diverse galaxies}
\label{sec:construct_fNHI_error1}
We quantify systematic uncertainties in $f(\NHI)$ that are caused by distance errors, stochastic sampling of galaxies in each $\lgNHI$ bin, and stochastic sampling of high galaxy inclinations.

\begin{figure}
  \centering
   \includegraphics{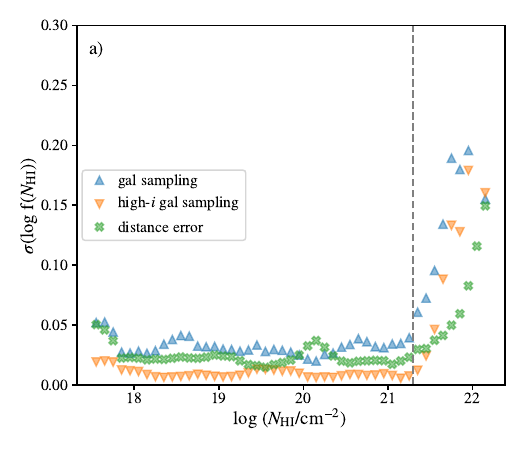}
    \includegraphics{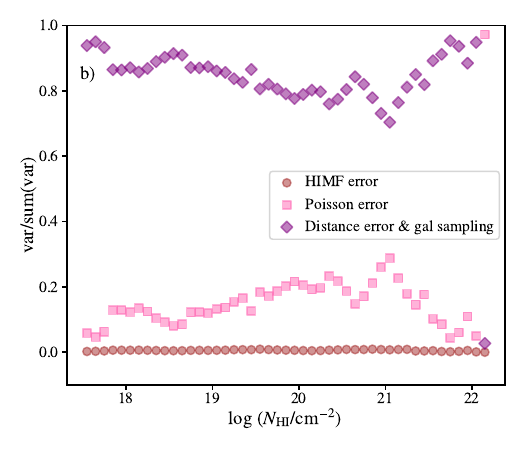}
  \caption{  Systematic error estimation for $f(\NHI)$. {\bf Panel a:} different errors as a function of $\lgNHI$. Three types of errors are considered, due to the stochastic sampling of galaxies (blue upward triangles), due to stochastic inclusion of highly inclined galaxies (orange downward triangles), and distance errors (green crosses). The vertical line marks $\lgNHI=21.3$, above which the errors increase dramatically . {\bf Panel b:} the contributions of different variance components to the final variance of $f(\NHI)$. Three types of variances are considered: HIMF parameter error of \citet{Guo23} (brown circles), Poisson error from the counts of $\lgNHI$ areas (pink squares), and the combined effects of distance errors and galaxy stochastic samplings (purple diamonds).
 }
  \label{fig:bootstrap}
\end{figure}

The distances of the main sample have a typical uncertainty of 30\% \citep{Kourkchi17,Leroy19}.
We quantify the $f(\NHI)$ uncertainty caused by this factor through randomly shifting the distances by an extent within $\pm30\%$.
We build 100 sets of such mock catalogs, repeat the $f(\NHI)$ construction procedure for each set, and derive the standard deviations between these 100 sets of $f(\NHI)$.
These standard deviations are taken as systematic uncertainties associated with distance errors (green crosses in Figure~\ref{fig:bootstrap}-a).
These uncertainties are around 0.02 dex when $\lgNHI<21.3$, and become higher at the high-$\NHI$ end.
The high-$\NHI$ areas tend to be in galaxy inner regions, are less self-similar or correlated in size with $\MHI$ than the outer $\hi$ disks \citep{Wang25}, and thus more sensitive to distances.

Quantifying the systematic uncertainty due to stochastic sampling of galaxies is motivated by the diverse $\hi$ morphology observed in and around galaxies \citep{deBlok24,Wang24}.
We derive the scatter of area counts as a function of $\NHI$ in each $\MHI$ bin that have more than six galaxies, and find values ranging from 0.2 to 0.4 dex.
Such systematic differences between individual galaxies can contribute significantly to final errors, when the galaxy sample is not sufficiently large.
We quantify the associated $f(\NHI)$ uncertainty through bootstrapping the galaxies in each $\MHI$ bin.
In each bootstrapping round, for each $\MHI$ bin with at least two galaxies, we resample the galaxies with replacement.
Only the $\MHI \sim 10^{8.5} \, M_{\odot}$ bin has one galaxy and is excluded from this procedure.
We repeat this procedure and produce 100 sets of resampled data, and derive $f(\NHI)$ for each set.
The standard deviations of $f(\NHI)$ among these 100 deviations as a function of $\NHI$ are the systematic uncertainties associated with stochastic sampling of galaxies in $\MHI$ bins (blue dots in Figure~\ref{fig:bootstrap}-a).
The typical uncertainties throughout the $\lgNHI$ range of 18 to 21.3 are $\sim$0.025 dex.
They rise steeply at higher $\NHI$ values where the area counts are small (section~\ref{sec:construct_fNHI_data}) and more prone to galaxy variations, reaching 0.25 dex at $\lgNHI=22$.

The systematic uncertainty due to stochastic sampling of highly inclined galaxies is motivated by the fact that the $\lgNHI>21.1$ areas primarily reflect projection effects.
The deprojected $\hi$ surface densities rarely exceed $10^{21.1}\,\cmsq$, above which $\hi$ is efficiently converted to molecular gas \citep{Bigiel08}.
We evaluate the inclination associated systematic effects through bootstrapping the sub-sample with axis ratio below 0.3, corresponding to $i>72\degree$.
These highly inclined galaxies are resampled with replacement, producing 100 data sets for calculating $f(\NHI)$ scatters.
The deviations (orange downward triangles in Figure~\ref{fig:bootstrap}-a) are only significant when $\lgNHI>21.5$, and there they are close to those associated with stochastic samplings of galaxies in each $\lgNHI$ bin (blue upward triangles).
This deviation pattern confirms that the $f(\NHI)$ at $\lgNHI>21.1$ is mostly shaped by the high-inclination galaxies, and the errors there are dominated by stochastic sampling of these galaxies.

When we put the $f(\NHI)$ errors together, Figure~\ref{fig:bootstrap}-b shows that, in the $\lgNHI$ range between 17.8 and 21.3, the stochastic sampling of galaxies and distance errors dominate and contribute $>80\%$ of the $f(\NHI)$ variance in this study.
The variance due to Poisson error of area counts contribute secondarily and only roughly 20\% to the final variance.
The contribution of HIMF errors are minimal in the final variance.

To sum up, when $17.8<\lgNHI<21.3$, $f(\NHI)$ is relatively accurately determined, with errors less than 0.05 dex.
When $\lgNHI>21.3$, $f(\NHI)$ is highly uncertain because the stochastic sampling is worsened by the even lower number of highly-inclined galaxies than the main sample.

\subsubsection{Systematic uncertainties associated with the full-HI constructing procedure}
\label{sec:construct_fNHI_error2}
\begin{figure}
  \centering
  \includegraphics{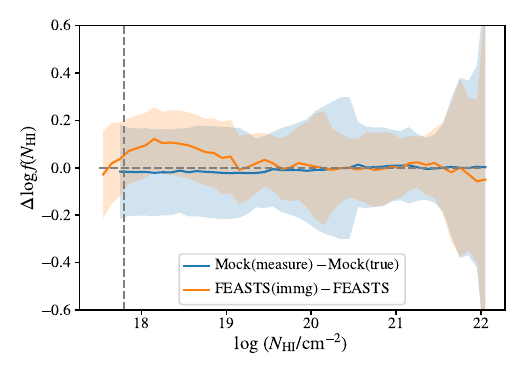}
  \caption{ Systematical shifts of $f(\NHI)$ due to different procedures of producing the full-$\hi$ images. Curves plot difference in $f(\NHI)$ between full-$\hi$ mock images produced with different procedures.
  The $f(\NHI)$ from images process with the fiducial procedure is compared with true answer: Mock(measure)-Mock(true).
  The $f(\NHI)$ from FEASTS images processed with the old procedure is compared with that of the new method: FST(immg)-FST.
  The error bars are the propagation of $f(\NHI)$ errors described in section~\ref{sec:contruct_fNHI}, and not the scatters of offsets.
  The vertical dashed line marks the median limiting $\lgNHI$ of 17.8.
  }
  \label{fig:fNHI_test1}
\end{figure}

We conduct mock tests to quantify possible errors related to the full-$\hi$ image construction (FAST+interferometry combination) for the FEASTS sample.
Procedures producing the mock single-dish and interferometric image data are introduced in Appendix~\ref{app:mock}.
These mock data are processed and analyzed in the same way as for the real data.
The mock measurements are compared with the true values as Mock(measure)$-$Mock(true) in Figure~\ref{fig:fNHI_test1}.
The median systematic differences between the two are close to zero, suggesting that the fiducial full-$\hi$ image construction does not bring significant systematic biases to the $f(\NHI)$.
We therefore do not consider this type of errors in the $f(\NHI)$ error calculation.

In Figure~\ref{fig:fNHI_test1}, we also evaluate the systematic differences in $f(\NHI)$ between the old and new procedures of constructing FEASTS full-$\hi$ images (section~\ref{sec:combine_fst}).
The $f(\NHI)$ from the old procedure (FST(immg)) shows relatively small differences ($<0.05$ dex) from that using the new and fiducial method (FST), when $\lgNHI>18$.
When $18<\lgNHI<18.8$, the differences are most apparent, approaching 0.1 dex.
This type of error with the old imaging method will not decrease when the galaxy samples increase in the future.
It is thus recommended to use the new procedure to build the full-$\hi$ image, to reduce the errors as much as possible.

\subsubsection{Systematic uncertainties associated with the optically thin assumption of HI}
\label{sec:construct_fNHI_abs}

 We have assumed the $\hi$ to be optically thin when converting the moment 0 images to column density maps.
As pointed out in Z05, the optically thin approximation may break down for the high-$\NHI$ ($>10^{21} \cmsq$) $\hi$, leading to an underestimate of $\NHI$ and artificial change in shape of $f\NHI)$ there (also see \citealt{Braun12}).
But the effect should be insignificant for the remaining intermediate and low-$\NHI$ regime, which is the main focus of this study.
We estimate the systematic uncertainty associated with the optically thin assumption using the test below.

\begin{figure}
  \centering
  \includegraphics{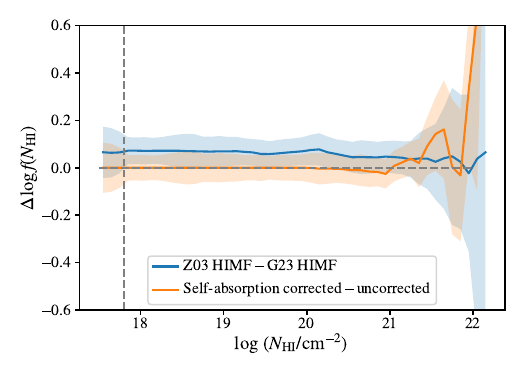}
  \caption{ Systematical uncertainties of $f(\NHI)$ related to 21-cm self-absorption and HIMF systematics.
  The blue curve plots the difference between assuming the \citet[][Z03]{Zwaan03} HIMF and the fiducial \citet[][G23]{Guo23} HIMF.
  The orange curve plots the difference between correcting and not correcting for self-absorption of the $\hi$ 21-cm emission line.
  The error bars (shaded areas) are the propagation of $f(\NHI)$ errors described in section~\ref{sec:contruct_fNHI}, and not the scatters of offsets.
  The vertical dashed line marks the median limiting $\lgNHI$ of 17.8.
  }
  \label{fig:fNHI_test_himf_abs}
\end{figure}

Previous studies have developed a ``sandwich'' model of the cool $\hi$ surrounded by two layers of warm $\hi$, to explain the tight relationship between $\hi$ opacity and apparent $\NHI$, based on 21-cm emission and absorption observations of the Milky Way \citep{Kanekar11, Braun12}.
Adopting this model (Eq. 8 and 9 from \citealt{Braun12}), we correct for the $\hi$ self-absorption when calculating the column densities.
The orange curve in Figure~\ref{fig:fNHI_test_himf_abs} shows that the difference of $\gtrsim$0.03 dex between the corrected and direct $\NHI$, which is found to be 0.03 dex at a direct $\lgNHI$ of 21, and leads to an offset of -0.025 dex in $f(\NHI)$ at $\lgNHI=21$.
The differences in $f(\NHI)$ between corrected and direct $\NHI$ quickly drop to zero toward lower $\lgNHI$, while they fluctuate around 0.1 with a large scatter due to poor statistics when $\lgNHI>21$.

We thus conclude that for the $\lgNHI<21$ regime, the $\hi$ self-absorption does not significantly affect our results.
The cool and warm $\hi$ partition is still poorly constrained in external galaxies (see \citealt{Pingel24}).
Recent studies show evidence that some early conclusions on significant 21-cm self-absorption originated from blended and confused line-of-sight profiles, and that kinematically there is no strong evidence for optically thick $\hi$ throughout galactic disks \citep{Koch21}.
Therefore, we do not correct for this systematic offset due to self-absorption or add this uncertainty to the error bars of $f(\NHI)$.
Finally, the integral properties like cross-section, length incidence (see Section~\ref{sec:discuss_incidence}), and covering fraction (see Section~\ref{sec:discuss_b}) for $\lgNHI>17.8$ should not be affected by the self-absorption.

\subsubsection{Systematic uncertainties associated with different HIMF}
\label{sec:construct_fNHI_HIMF}
The \citet{Guo23} HIMF has been used because of its lower parameter error than the \citet{Zwaan03} HIMF used in Z05.
There is a systematic difference between HIMFs obtained from different datasets, reflecting systematic uncertainties because of data incompleteness and systematic dependence because of cosmic variance \citep{Zwaan05b, Ma25}.
Specifically, \citet{Ma25} showed that there is 0.2-0.3 dex difference in the high mass end of HIMF ($\log \MHI/M_{\odot}\geq 10$) between \citet[][updated HIMF based on \citet{Zwaan03}]{Zwaan05b} and \citet{Guo23}.
Given that 13 out of 70 galaxies in the main sample have $\log \MHI/M_{\odot}\ge 10$, the systematic difference in HIMF at the high-mass end may contribute to the systematic errors in $f(\NHI)$.

We evaluate this HIMF related uncertainty through deriving $f(\NHI)$ with the \citet{Zwaan03} HIMF, and compare the result to the fiducial one derived with the \citet{Guo23} HIMF.
The blue curve in Figure~\ref{fig:fNHI_test_himf_abs} shows that there is an almost constant excess of 0.075 dex throughout the $\NHI$ range in the former.
Therefore, any offset of $f(\NHI)$ between populations should be larger than 0.075 dex in order to be concluded physically robust against systematic uncertainties due to HIMF differences.

\subsection{Additional systematic dependence} 
\label{sec:result_systematics}
As a final step before presenting the derived $f(\NHI)$, we discuss additional factors contributing to systematic shifts of $f(\NHI)$, including different sample or data (FEASTS versus MHONGOOSE), and different spatial resolution.
These two effects characterize possibly intrinsic dependence of the $f(\NHI)$, but are not taken as errors.

For these quantifications of systematics, we derive a 0.3-kpc resolution $f(\NHI)$ by constructing full-resolution MHONGOOSE images that start from the r00\_t00 type (in contrast to the fiducial 1-kpc resolution measurements that are based on MHONGOOSE images starting from the r10\_t00 type).
We additionally construct 10-kpc resolution $f(\NHI)$ for the main sample, using the deblended and cleaned FEASTS images without combining them with the interferometry data, and smoothing the MHONGOOSE r10\_t90 images to the 3.24$'$ angular resolution of FEASTS.

\begin{figure}
  \centering
  \includegraphics{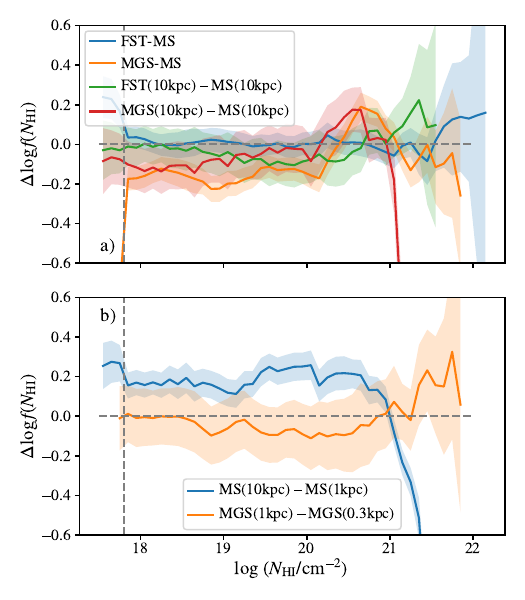}
  \caption{ Systematical shifts of $f(\NHI)$ due to different samples and resolutions. Curves plot difference in $f(\NHI)$ between different datasets, with details in section~\ref{sec:result_systematics}.
    The vertical dashed lines mark the median limiting $\lgNHI$ of 17.8.
  {\bf Panel a:} differences in $f(\NHI)$ between different samples with the same resolution. The main sample (MS), the FEASTS sample (FST), and the MHONGOOSE sample (MGS) are compared. In addition to the fiducial 1-kpc resolution (if not specified), $f(\NHI)$ are also derived and compared at the 10-kpc resolution (denoted in the figure labels).
  {\bf Panel b:} differences in $f(\NHI)$ between the same sample with different resolutions. Three sets of resolutions are considered: 0.3 kpc (the MGS only), 1 kpc and 10 kpc.
 }
  \label{fig:fNHI_test2}
\end{figure}

\subsubsection{Sample dependence}
For the systematics related to sample biases, we derive $f(\NHI)$ for the MHONGOOSE and FEASTS sub-samples separately, and subtract from them the main-sample $f(\NHI)$ so as to zoom in on the differences.
Before looking into the difference, we note that, neither FEASTS nor MHONGOOSE have sufficient statistics to fill all the $\MHI$ bins (Figure~\ref{fig:full1kpchist}-e);
it means that both $f(\NHI)$ are lower limits due to the methodology of weighting galaxies by the HIMF $\psi(\MHI)$.
It also indicates that the main sample $f(\NHI)$ does not necessarily lie between these two curves in all $\NHI$ bins.
So the difference quantified here should be viewed as upper limits to systematic differences due to sample biases or data differences.

In Figure~\ref{fig:fNHI_test2}-a, the FEASTS $f(\NHI)$ is systematically higher than the MHONGOOSE $f(\NHI)$ by 0.1 to 0.2 dex when $\lgNHI<19.8$.
A similar extent of differences is found when comparing the $f(\NHI)$ measurements from the two datasets at a resolution of 10 kpc for $\lgNHI<19$, implying that the former difference is not totally due to detailed difference in data capabilities of resolving intermediate-scale diffuse structures.
At least part of the difference is likely a true sample difference, possibly related to different $\hi$ structures around low- (MHONGOOSE) and high-mass (FEASTS) galaxies, but the extent should on average be lower than 0.2 dex after removing possible upper-limit effects, which should be better quantified in the future after enlarging the samples.

\subsubsection{Resolution dependence}
Figure~\ref{fig:fNHI_test2}-b shows that the MHONGOOSE 0.3-kpc result differs little from the MHONGOOSE (1-kpc) one considering the error bars.
The 1-kpc version only has a very small difference with respect to the error bars from the 0.3-kpc version when $18.8<\lgNHI<21$.
It indicates that at the 1-kpc resolution, the $f(\NHI)$ can be relatively robustly derived for LLS-like systems.

In Figure~\ref{fig:fNHI_test2}-b, we find that the $f(\NHI)$ of the main sample at the 10-kpc resolution is overestimated compared to that at the 1-kpc resolution by an almost constant excess of $\sim$0.2 dex when $\lgNHI<20.8$, compensated by significant under-estimates at higher $\NHI$.
It suggests that between the 1-kpc and 10-kpc resolution, image smoothing tends to shift the $f(\NHI)$ near the LLS regime upward in an almost parallel way.
The shifting extent is not large compared to 1-kpc $f(\NHI)$ error bars or MHONGOOSE versus FEASTS differences, both urging the need to increase 1-kpc-resolution sample size, and pointing to the potential of future lower-resolution observations in revealing distribution structures for the $\lgNHI\sim18$ gas.

With the MHONGOOSE sample, we find a tentative contrary trend when increasing the 0.3-kpc resolution to the fiducial 1 kpc: some $\hi$ looks to be shifted from $18.5<\lgNHI<21$ to $\lgNHI>21$.
Because the $\lgNHI>22$ $\hi$ in the 0.3-kpc images is smoothed out into the $21<\lgNHI<22$ range and disappears at the 1-kpc resolution.

These tests suggest that $f(\NHI)$'s derived from different resolutions can have systematic differences of 0.1 to 0.2 dex, and the offset at a given $\NHI$ is not necessarily a monotonic function of resolution, possibly a consequence of the intrinsic $\hi$ hierarchical structure.
When the $f(\NHI)$ is compared to that of the pencil-beam LLS observations, we should keep such systematic differences in mind.

\section{Result: the column density distribution } 
\label{sec:result}
\begin{figure*}
  \centering
  \includegraphics{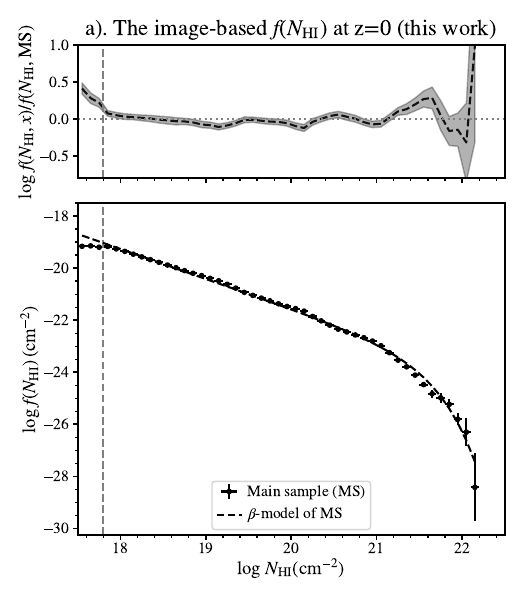}
    \includegraphics{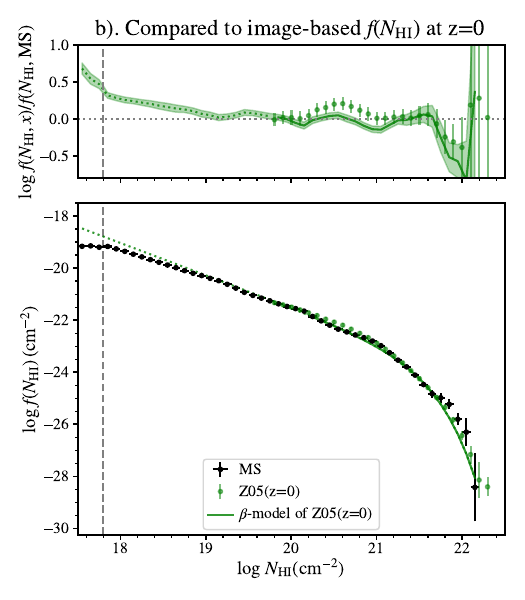}
      \includegraphics{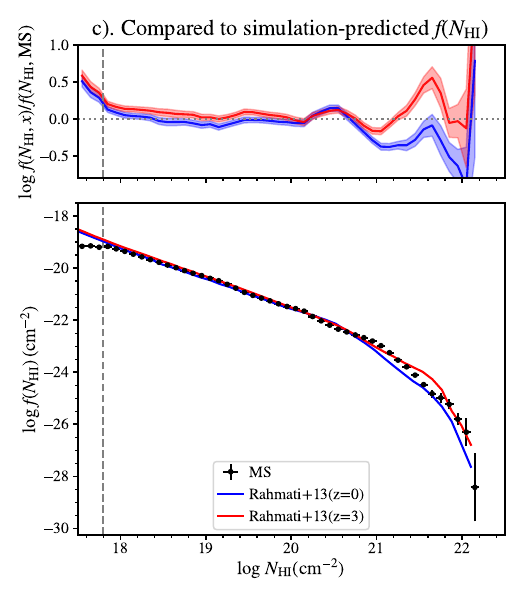}
        \includegraphics{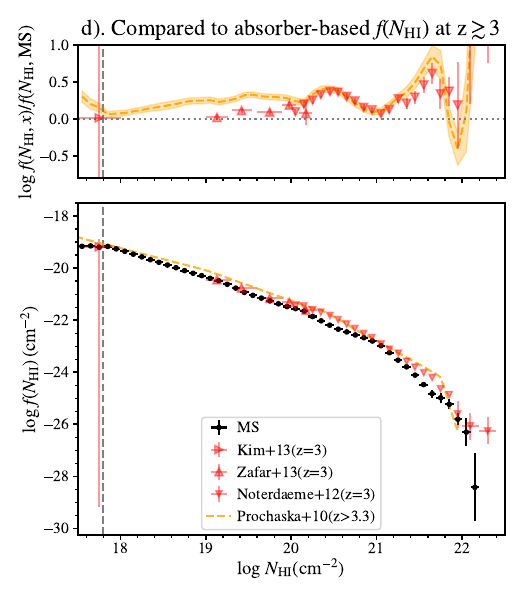}
  \caption{ The $\hi$ column density distribution function $f(\NHI)$ to the level of $10^{17.8}\,\cmsq$ for $z=0$, and its comparison to literature results.
  In each set, the lower panel shows the $f(\NHI)$, and the upper panel the difference of best-fit models or literature results to the $f(\NHI)$ derived in this study; the vertical dashed line mark the median $N_\mathrm{HI,lim}$ of the main sample.
  {\bf Panel a:} the direct derivation of $f(\NHI)$ based on the main sample (MS) in this study is plotted as black dots and error bars. Its best-fit Schechter function is plotted as the purple curve.
  {\bf Panel b:} comparing to the $f(\NHI)$ of Z05, which is also derived from 21-cm emission line images but for $\lgNHI>19.8$. The direct data points of Z05 are plotted as green dots, and its best-fit Schechter function is plotted as green solid and dotted curves when $\lgNHI>19.8$ and $<19.8$ respectively.
  {\bf Panel c:} comparing to the simulation prediction from \citet[][OWLS]{Rahmati13} for $z=0$ (blue curve), and $z=3$ (red curve).
  {\bf Panel d:} comparing to absorber-based $f(\NHI)$ at higher redshifts. The absorption based measure of \citet[][UVES]{Kim13}, \citet[][UVES]{Zafar13}, and \citet[][SDSS]{Noterdaeme12} for $z\sim3$, are plotted in red rightward, upward, and downward triangles respectively. The absorption based piece-wise power law model of \citet[][fit to SDSS observation]{Prochaska10} for $z>3.3$ is plotted as a orange dashed curve.
 }
  \label{fig:fNHI}
\end{figure*}

Figure~\ref{fig:fNHI}-a shows the main result of this study, the $f(\NHI)$ constructed from the main sample.
The dashed vertical line marks the 50\%-complete $\NHI$ limit, below which $f(\NHI)$ is less reliable.

Following the literature including Z05, we fit a Schechter function \citep{Schechter76} to the $f(\NHI)$ for the $\lgNHI$ range of 17.8 to 22:
\begin{equation}
f(\NHI)=\frac{f^*}{N^*_\mathrm{HI}} ( \frac{\NHI}{\NHI^*} )^{-\beta} e^{-\NHI/\NHI^*}
\label{eq:fNHI}
\end{equation}
The model thus consists of a power-law part for the low-$\lgNHI$ regime with a logarithm slope of $-\beta$, and an exponentially declining part for the high-$\lgNHI$ regime with a characteristic column density of $\NHI^*$.
The Schechter function is commonly used to characterize galactic luminosity functions and mass functions \citep[e.g.,][]{Zwaan03, Guo23}.

The fitting is conducted in the logarithm space including the resolution-limited and uncertain $\lgNHI>21.5$ side, mostly for capturing the total incidence of $\lgNHI>18$.
The likelihood function is assumed to be a simple Gaussian where the variance is underestimated by a fractional amount, $f$.
The best-fit parameters (numerical optimum of likelihood function) are obtained with the scipy.optimize module.
The posterior probability function and errors are estimated with the emcee \citep{ForemanMackey13} algorithm.
Uniform distributions have been assumed, and ranges are set to be $(20.8\, , 22.6)$, $(0.8\,, 1.5)$, $(-2.5\,,-1)$ respectively for priors on $\log \NHI^*$, $\beta$,  $\log f^*$, and $\log f$.

The best-fit parameters and errors are listed in Table~\ref{tab:fNHI_para}.
The standard deviation of data points around the best-fit model is 0.067 dex, and the data variance underestimation is lower than 0.01\%.
This model and the deviation from data points are also displayed in Figure~\ref{fig:fNHI}-a.
The emcee \citep{ForemanMackey13} generated corner figure can be found in Figure~\ref{fig:mcmc} in Appendix~\ref{app:fNHI}.
The figure shows that most projected posterior probability functions center on the best-fit values except for a negligible underestimate of $\log f$, and the covariance between parameters are relatively weak except for that between $\beta$ and $\log f^*$.
We do not show the parameter uncertainty associated model curve broadening, because it is very narrow.

{ \color{green}
\begin{table}
  \centering
          \caption{Parameters of the best-fit Schechter function} of $f(\NHI)$.
          \begin{tabular}{c c c  c}
            \hline
            $\log (\NHI^*/\mathrm{cm}^{-2})$ &  $\beta$ &  $\log f^*$  & $\log f$ \\
            \hline
           $21.26^{+0.02}_{-0.02}$ & $1.150^{+0.016}_{-0.016}$ & $1.754^{+0.036}_{-0.037}$ &  $-5.491^{0.128}_{-0.124}$ \\
            \hline
          \end{tabular}

      \label{tab:fNHI_para}
  \end{table}
}

When $\lgNHI<17.8$, the measurement flattens compared to the best-fit model, most likely reflecting detection limit effects.

We compare our $f(\NHI)$ results to the direct measurements and best-fit Schechter function of Z05 at $z=0$ (Figure~\ref{fig:fNHI}-b), which are based on data with a higher sensitivity limit of $\lgNHI>19.8$ but similar spatial resolution as our data.
Our $f(\NHI)$ measurements are comparable with the Z05 Schechter function in the $19.0<\lgNHI<21.6$ range, but are $\sim0.2$ dex lower than the Z05 measurements in the $20.2<\lgNHI<20.8$ range.
Considering the HIMF difference (section~\ref{sec:construct_fNHI_HIMF}), these two $f(\NHI)$ measurements should be more consistent with each other if using the same HIMF.
Therefore, in the $19.8< \lgNHI<21.6$ range, our $f(\NHI)$ measurements are consistent with the Z05 results.

Our $f(\NHI)$ measurements deviate from the Z05 extrapolation at low $\NHI$ values.
The deviation increases to around 0.3 dex at $\lgNHI=18$, reflecting possible error amplification in extrapolation.
At the highest column densities of $\lgNHI>21.3$, our measurements also deviate from the Z05 curve, possibly due to high systematic uncertainties.
Although the Z05 sample has a typical resolution ($\sim1.3\,$kpc for their fiducial highest-angular-resolution dataset) similar to the main sample here ($\sim1.2\,$kpc), small difference in the resolution distribution shape may also add to the $f(\NHI)$ difference shown here.

In Figure~\ref{fig:fNHI}-c, the $f(\NHI)$ is also compared to the prediction of the cosmological simulation OWLS (Overwhelmingly Large Simulations, \citealt{Schaye10}) in \citet{Rahmati13} (see also \citealt{Rahmati15} for the Evolution and Assembly of Galaxies and their Environment, EAGLE, simulation, \citealt{Schaye15}) for $z=0$.
In that simulation, radiative transfer was performed in post-processing to account for ionizing UVB radiation and ionizing recombination radiation \citep{Rahmati13b}.
The \citet{Rahmati13} prediction roughly matches the $f(\NHI)$ with offsets fluctuating around zero within the error bars when $\log\NHI<20.2$.
Particularly, the prediction seems to match perfectly the best-fit Schechter function in the same $\NHI$ range, as can be seen from the similarity in residual shapes with respect to the direct measurements (top panels of Figure~\ref{fig:fNHI}-c and \ref{fig:fNHI}-a).
At higher $\NHI$, the \citet{Rahmati13} prediction can be less accurate, as pointed out by the authors, because ISM physics were treated in a relatively simple way there.

Studies using simulations to predict $f(\NHI)$ or $\hi$ density distribution for the $\lgNHI>17.2$ regime and compare to observations have so far concentrated on the $z\sim3$ slice of the Universe \citep{Rahmati13, Rahmati15, VillaescusaNavarro18, Sokoliuk25}.
Previous simulation-observation comparisons for $z=0$ instead focus more on reproducing the integral $\hi$ mass function and scaling relations, or distribution of $\lgNHI\gtrsim20$ \citep{Crain17, Diemer18, Dave20}.
There are also simulation studies comparing $f(\NHI)$ $z=0$ predictions between recipes for the IGM at much lower $\NHI$ or the ISM at the high-$\NHI$ end \citep[e.g.,][]{Crain17, Tillman23}.
Now it is possible to compare simulation and observation for the $\lgNHI>17.8$ $\hi$ distribution.
The observational $f(\NHI)$ derived here for $z=0$ (and related quantification in section~\ref{sec:discuss_b} and section~\ref{sec:discuss_incidence}) provides a new anchor point for future simulation work.

\section{Discussion: relation with LLS column density functions, impact parameters and incidence} 
\label{sec:discuss}
\subsection{A modest decline of $f(\NHI)$ since $z=3$}
We overplot in Figure~\ref{fig:fNHI}-d, the $f(\NHI)$ compiled by \citet{Zafar13}, with LLS, sub-DLA, and DLA observations at $z=3$.
They are based on data of SDSS (Sloan Digital Sky Survey) by \citet{Noterdaeme12} , UVES (Ultraviolet Visual Echelle Spectrograph) by \citet{Kim13}, and UVES by \citet{Zafar13}, respectively.
The SDSS based result for $z>3.3$ from \citet{Prochaska10} is also plotted in Figure~\ref{fig:fNHI}-d, which consists of 4 power law segments (thus more parameters than the Schechter function) fitted to their data in the $\NHI$ range plotted here.
We recall that the comparison is most robust for $\lgNHI<21$, where the optically thin assumption works well (section~\ref{sec:construct_fNHI_abs}).

The $z>3.3$ $f(\NHI)$ is almost always higher than the $z=0$ one by $\gtrsim0.2$ dex.
For both $z>3.3$ and $z=3$, the large excess in $f(\NHI)$ of $\lgNHI>21$ should be lowered by a median of $\sim$0.1 dex, if self-absorption is corrected in the $z=0$ measurements.
Systematics due to small sample size and self-absorption of the $z=0$ $f(\NHI)$ possibly produce the large fluctuation in $f(\NHI)$ difference at $\lgNHI>21$.

The $z=3$ absorption based $f(\NHI)$ is systematically higher than the $f(\NHI)$ at $z=0$ by an offset of 0.2 to 0.4 dex when $19.2<\lgNHI<21$.
The differences become smaller toward lower $\lgNHI$, with offsets 0.1--0.2 dex for $19.2<\lgNHI<20$, slightly larger than systematic uncertainties related to the HIMF variance.
There is a hint for the offsets to disappear ($<$0.1 dex) when $17.8<\lgNHI<19.2$, but the LLS sampling in $\lgNHI$ also becomes much sparser.

The $z=0$ result is at a resolution of around 1 kpc, while the covering fraction of $\hi$ clouds in a 1-kpc pixel, $f_\mathrm{cloud}$, should be $\leq1$.
So the actual relative difference of the $z\gtrsim3$ absorption-based measurements from the $z=0$ ones needs be multiplied by a factor of $1/f_\mathrm{cloud}$ (i.e., the offsets increasing by 0.16 dex if $f_\mathrm{cloud}=0.70$, more on this in section~\ref{sec:discuss_incidence}).
Therefore, we conclude that it is possible for the $f(\NHI)$ of LLS-like column densities to have declined slightly from $z=3$ to $z=0$.
Such a modest evolution was not predicted in the previous simulation OWLS \citep{Rahmati13} as shown in Figure~\ref{fig:fNHI}-c.

Although the UVB has declined by more than an order of magnitude since $z = 3$, previous studies have found that the evolution of $f(N_{\mathrm{HI}})$ remains relatively modest for LLS, in contrast to the $3<z<6$ Universe where the LLS incidence drops quickly with decreasing redshift \citep{Ribaudo11, Fumagalli13}.
This relatively modest evolution reflects a balance among several competing processes: the weaker UVB intensity tends to increase the neutral fraction, whereas collisional ionization in the increasingly hot CGM and the gradual decrease in the cosmic mean density act in the opposite direction.
Feedback from star formation and active galactic nuclei further contributes by heating halo gas and enhancing collisional ionization, thereby lowering the neutral fraction and suppressing high-$N_{\mathrm{HI}}$ systems.

The partial cancellation among these radiative, thermal, and dynamical effects, together with continued structure growth and self-shielding in dense regions, helps preserve a broadly similar shape of $f(N_{\mathrm{HI}})$ from $z \sim 3$ to the present.
The systematic difference we observe between $z = 0$ and $z = 3$ thus captures the net outcome of these intertwined processes.

\begin{figure}
  \centering
  \includegraphics{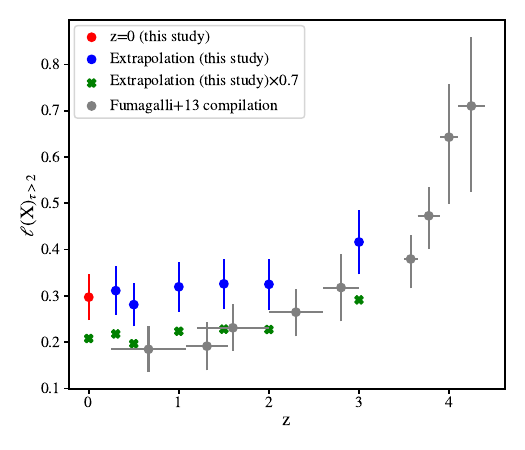}
  \caption{  Length incidence for $\lgNHI>17.5$ (optical depth $\tau>2$). The red and blue dots are calculations based on $z=0$ observations in this study. The direct $z=0$ calculation (red dots) is based on cumulating the best-fit Schechter function of this study. The extrapolations (blue dots) are based on $\hi$ images in the main sample at $z=0$, but HIMFs are the NUM predictions at different redshifts \citep{Guo23}.
  The green 'X' symbols are the $z=0$ direct and extrapolated calculations times a factor of 0.7 to guide the eye.
  The grey dots are LLS observations at different redshifts compiled by \citet{Fumagalli13}, from studies of \citet{Prochaska10, Fumagalli13, OMeara13, Ribaudo11} for the redshifts range $z>3.5$, $z\sim3$, $1.5<z<2.5$, and $0.5<z<1.5$, respectively.
 }
  \label{fig:dNdX}
\end{figure}

\subsection{The length incidence}
\label{sec:discuss_incidence}
We relate the $f(\NHI)$ measurements to length incidence (or line-of-sight density) of LLS absorbers at different redshifts.

For such comparison, we calculate the length incidence of, or reciprocal of mean free path between, absorbers with column density greater than $\NHI$, by
\begin{equation}
\ell(N_\mathrm{HI})=\frac{ \mathrm{d}N(N_\mathrm{HI})}{\mathrm{d}X}.
\end{equation}
Here the number density, $\mathrm{d}\, N(N_\mathrm{HI})/\mathrm{d}X$ for column densities larger than $N_\mathrm{HI}$, is derived through integrating the $f(\NHI)$ measurements or Schechter function: $\mathrm{d} N(N_\mathrm{HI})/\mathrm{d} X=\int^{10^{22.5}}_{N_\mathrm{HI}} f(\NHI') \, \mathrm{d} \NHI' $.
It is related to the redshift number density $\mathrm{d}\, N(N_\mathrm{HI})/\mathrm{d}z$ through:
$\frac{ \mathrm{d}N(N_\mathrm{HI})}{\mathrm{d}X}=\frac{\mathrm{d} N(N_\mathrm{HI})}{\mathrm{d}z}\frac{\mathrm{d}z}{\mathrm{d}X}$,
where $\mathrm{d}X=\frac{H_0}{H(z)}(1+z)^2 \mathrm{d}z$.
At $z=0$, the $\ell(N_\mathrm{HI})=\mathrm{d}\, N(N_\mathrm{HI})/\mathrm{d}z$, as $\mathrm{d}z/\mathrm{d}X=1$.

Through integrating the directly measured $f(\NHI)$ curve, we obtain $\ell(10^{17.8}\,\cmsq)=0.267\pm0.033$.
To better match the LLS column density range, we integrate the best-fit Schechter function, and obtain $\ell(10^{17.5}\,\cmsq)=0.298\pm0.045$, where $\lgNHI\ge 17.5$ corresponds to an optical depth at the Lyman limit of $\tau\ge 2$.
 Through a similar calculation, we obtain $\ell(10^{17.2}\,\cmsq)=0.345\pm0.052$, and $\ell(10^{17.8}\,\cmsq)=0.256\pm0.039$.
\citet{Fumagalli13} pointed out that the varying results on $\ell$ in early studies could be partly due to inconsistent definition of minimum $N_\mathrm{HI}$ in the integration \citep{Lanzetta91,StorrieLombardi94,Songaila10}.
Our calculation above indicates an increase in $\ell(\NHI)$ by a factor of 1.35 at $z=0$ when the minimum $N_\mathrm{HI}$ decreases from $10^{17.8}$ to $10^{17.2}\,\cmsq$.

For reference, we also derive $\ell(10^{20.3}\,\cmsq)$ for DLA-like systems, but warn that our sample size is not optimized for it.
We obtain $0.037\pm0.002$ and $0.036\pm0.002$ with direct measurements and Schechter function respectively, both slightly lower than but roughly consistent with the value of $0.045\pm0.006$ derived in Z05.

\subsubsection{The redshift dependence of LLS-like length incidence }
The incidence rate can be understood as $\langle n\sigma \rangle$, i.e.,
the average of galaxy number density $n$ times the $\hi$ cross section $\sigma$, where $n$ is inferred by the $\psi$ weighting in our $f(\NHI)$ construction.
At a resolution of 1 kpc, the incidence rate can be further expressed as 1-kpc resolution smoothed $\langle n\sigma \rangle_{1-kpc}$ times the surface cloud covering fraction $f_\mathrm{cloud}$, where $f_\mathrm{cloud}$ is related to the 3D $\hi$ clumpiness, which is often described by the volume filling factor or clumping factor, and possibly linked to the turbulence \citep{Suresh19, Gronke20, Qu26}.

We can thus infer the evolution of LLS cross-section or covering fraction near $z=0$ with the following experiment.
We extrapolate the $f(\NHI)$ at $z=0$ to higher redshift by assuming that the $\hi$ distribution of each galaxy (equivalently, the $\NHI$ distribution at a given $\MHI$) remains the same and only HIMF changes as a function of $z$.
The HIMF of each z is predicted by the Neutral\-Universe\-Machine (NUM, \citealt{Guo23}).
We calculate the corresponding $\ell$ values of $\lgNHI>17.5$, $\ell(10^{17.5}\,\cmsq)=\ell_{\tau>2}$, and compare them to real LLS measurements in Figure~\ref{fig:dNdX}.
The extrapolated $\ell$ values are consistently higher than those at $0.5<z<3$ by a factor of 1.481$\pm$0.212.

This calculation of length incidence has extrapolated the Schechter function below the $\log N_\mathrm{HI,lim}$ of 17.8 to 17.5, and there is a possibility of $f(\NHI)$ flattening with respect to the Schechter function there.
However, even if we force a flattening of the $f(\NHI)$ to start at $\lgNHI=17.8$, it only reduces the gap between the $z$-extrapolated and really observed $\ell$ at $z>0.5$ by a factor of 16\%.
It is possible that NUM HIMF consistently over-estimates the $\hi$ galaxy densities at $z>0$ by 1.481 times, but the $\Omega_\mathrm{HI}$ through integrating the HIMF predicted by NUM matches the (Ly$\alpha$-based) observations at different redshifts \citep{Guo23}.

If the HIMF predictions of NUM is correct, then, either the $z=0$ LLS-like $\hi$ has its $\langle n\sigma \rangle$ increased by a factor of $1.481\pm0.212$ compared to $0.5<z<3$, or the typical $f_\mathrm{cloud}$ is close to $0.675\pm0.096$.
While a larger $\langle n\sigma \rangle$ at $z=0$ is possible as the UVB has declined by nearly 10 times since $z=1$ \citep{FaucherGiguere20}, the $f_\mathrm{cloud}=0.675\pm0.096$ of $\hi$ may be more likely, as recently studies on the low-temperature gas distribution and velocity structures suggest high level of turbulence \citep{Chen23}.

The implied $f_\mathrm{cloud}$ seems comparable to the covering fraction of $\sim$0.7 for strong \ion{Mg}{2} absorbers at $z < 0.5$ out to 75 kpc around MW-like galaxies \citep{Chen10} (also see \citealt{Kacprzak08, Barton09}), or possibly larger than that of $f_\mathrm{cov}\sim$0.1-0.65 for LLSs observed at $z\sim3$ \citep{Rudie12, Prochaska13a, Prochaska13b, Fumagalli14}.
For the comparison with $z\sim0$ ion absorbers, we note the possible difference in metallicities and hence possibly different structures.
For a broader comparison with $f_\mathrm{cov}$ of LLSs in the literature, when the LLS searching radius (often the virial radius) is larger than where most LLSs locate, we expect $f_\mathrm{cloud}>f_\mathrm{cov}$.

\begin{figure*}
  \centering
  \includegraphics{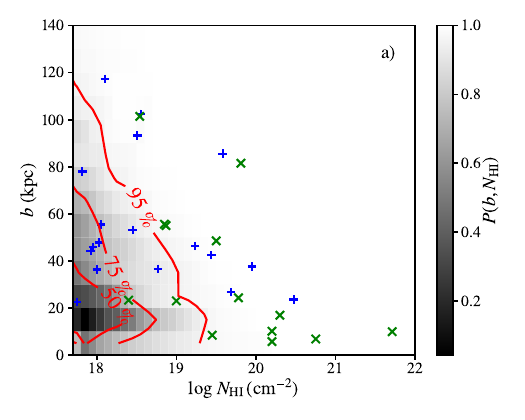}
  \includegraphics{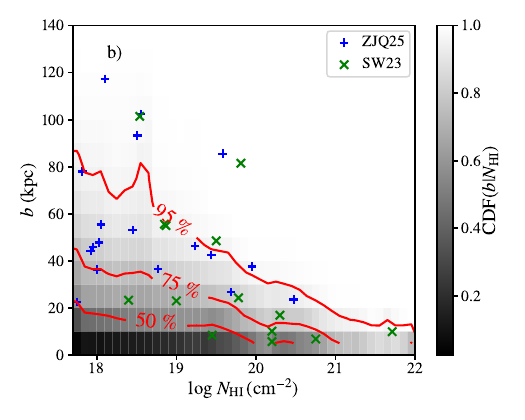}
  \caption{Probabilities of detecting $\hi$ in the space of $\NHI$ and impact parameter $b$.
   {\bf Panel a:} the colored map and associated contours show 2D cumulative probabilities of detecting $\hi$ in the space of impact parameter $b$ versus $\lgNHI$.
   {\bf Panel b:} the colored map and associated contours show cumulative conditional probabilities of detecting $\hi$ below $b$ as a function of $\NHI$.
    The probabilities are evaluated along different $b$ for a given $\NHI$, and the accumulation starts from the smallest toward the largest $b$ values}.
  In both panel a and $b$, the incidence of $z<0$ LLSs from two low-$z$ samples, \citet{Weng23} (green x) and \citet{Qu26} (blue $+$), are plotted for comparison.
  \label{fig:Pb}
\end{figure*}

\subsection{Probability of detecting HI as a function of $b$ and $\NHI$ }
\label{sec:discuss_b}
We investigate the expansion of $\hi$ as a function of column density limits around associated galaxies.
We derive cumulative probability distributions of impact parameter (or transverse distance) $b$ around the galaxy center for $\hi$ with different column densities, P($b$).
The P($b$) is constructed by weighting galaxy areas in the main sample $\hi$ images in a similar way as $f(\NHI)$, but is cumulated and normalized to a total probability of unity in the parameter space.
We compare P($b$) to distribution of data points from previous LLS observations at $z<0.5$.
The $z<0.5$ LLS observations are from two data collections: one is from the MUSE-ALMA survey \citep{Peroux22} with the $b$ derived in \citet{Weng23}, and the other is from \citet{Qu26} (Z. Qu in prep).
In both datasets, the lowest $b$ (of the nearest possible host galaxy) is taken for the analysis here, when there is confusion for an absorber to be associated with multiple possible galaxies.
So by construction, the LLS $b$ would be biased toward low values, if the galaxy catalog were complete.

We firstly investigate the relative probabilities of detecting $\hi$ for a given $b$ and $\NHI$.
Figure~\ref{fig:Pb}-a shows the cumulative probability distribution of $\hi$ in the $b$ versus $\lgNHI$ plane.
The accumulation starts from the highest probability density to the lowest one, so that 75\% of $\hi$ in the plotted $b$ versus $\lgNHI$ range should be detected within the 0.75-level contour.
A similar figure for DLA analogs can be found in Z05.
In Figure~\ref{fig:Pb}-a, 95\% of $\hi$ in the main sample is found with $\lgNHI<19.2$, but less than one third LLSs are found there.
The discrepancy implies that the absorber surveys of $\lgNHI>17.8$ at low redshifts are biased toward DLA-like ones.

We then investigate the probabilities of locating the $\hi$ at transverse distances below $b$ for a given detected $\NHI$.
The probabilities of $b$ are evaluated at a given $\NHI$, and accumulated starting from the smallest toward the largest $b$.
So that 75\% of the $\hi$ at a given $\lgNHI$ should be detected at $b$ lower than the 0.75-level line.
Figure~\ref{fig:Pb}-b shows this cumulative conditional probability distribution of $b$ given $\NHI$ as a function of $\NHI$; a similar figure for DLA analogs was presented in Z05.
There is considerable discrepancy in Figure~\ref{fig:Pb}-b between 21-cm and LLS observations.
The most striking one is that only 1 out of the 32 LLSs (+DLAs) are found within the 50\% contour of cumulative $b$ probability as a function of $\NHI$.
The discrepancy is too large to be explained with the possible $f_\mathrm{cloud}<1$.
Even if we assume an extremely small 10\% covering fraction, and shift the LLS data points left-ward by 1 dex, only 4 out of 32 LLSs (+DLAs) would lie below the 50\% cumulative probability curve.

This discrepancy could come from challenges in the LLS-galaxy pair identification, as pointed out in several recent studies \citep{Chen20, Weng23}.
Absorber studies have difficulty in associating absorbers to galaxies because there can be satellites, other haloes along the line of sight or even denser gas from the IGM contaminating the association.
A previous search in the TNG50 \citep{Nelson19,Pillepich19} simulation found that the actual contribution to the $\hi$ from the central galaxy diminishes very quickly with increasing impact parameter \citep{Weng24}.
Observationally, there is evidence that these stronger absorbers tend to be found in galaxy over-densities \citep{Peroux22, Qu23}, implying that they are probing gas in more large-scale environments than the central galaxies or localized galaxy associations.

\begin{figure*}
  \centering
    \includegraphics{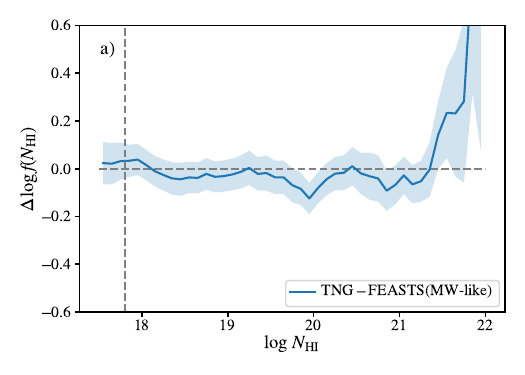}

  \includegraphics{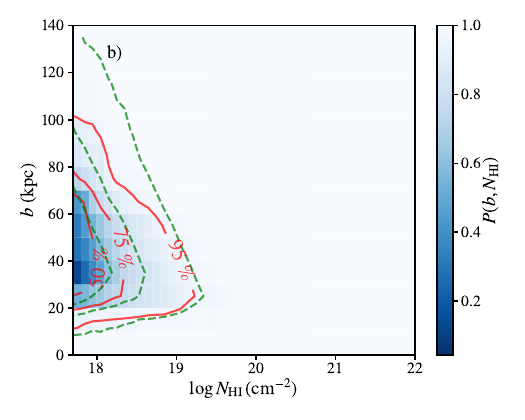}
  \includegraphics{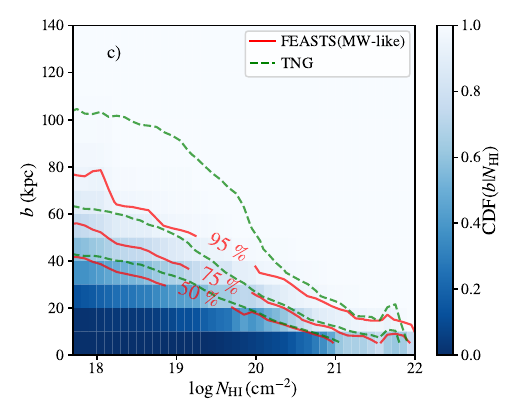}
  \caption{
    {\bf Panel a:} the offsets in $f(\NHI)$ between TNG50 predictions and observations for MW-like galaxies. The MW-like galaxies are selected from the FEASTS sample, and the TNG50 sample is a property matched control sample for the FEASTS sample \citep{Lin25b}.
    {\bf Panels b and c:} probabilities of $\hi$ detections in the 2D space of $\NHI$ and impact parameter $b$. Similar to Figure~\ref{fig:Pb}, but comparing the observation and the TNG50 simulation, and only for MW-like galaxies from FEASTS (red solid contours) and their control sample from the TNG50 (green dashed contours). }
  \label{fig:Pb_TNG}
\end{figure*}

\subsubsection{Comparing to the TNG50 predictions}
We briefly compare the $f(\NHI)$ and $b$ probability distribution to TNG50 predictions for a subset of MW-like galaxies.
\citet{Lin25b} conducted $\hi$ mock observations for 330 galaxies in the TNG50 simulation \citep{Nelson19,Pillepich19} that are matched in $M_*$, $\MHI$, inclination, and distance to a sub-set of 33 Milky Way like galaxies selected by $M_*>10^{9.5}\,M_{\odot}$ from FEASTS.
The $\hi$ postprocessing has been performed following contemporary standard procedures.
The neutral gas fraction is determined differently in star-forming and non-star-forming cells, following the procedure of \citet[][see also \citealt{Stevens19}]{Diemer18}.
The $\hi$-$\mathrm{H}_2$ partition follows the procedure of \citet{Marinacci17}, using the model of \citet{Leroy08}.
The total $\hi$ masses obtained for mock galaxies are consistent with the TNG50 catalog of \citet{Diemer18} with an offset of 0.038$\pm$0.028 dex.

The mock images are deblended with the same procedure as for the observational data of this paper.
There are 23 of the 33 galaxies in that sub-set included in the main sample.
We compare the measurements from the property-matched TNG50 mocks at a resolution of $\sim$1-kpc matched to these 23 FEASTS galaxies.

Figure~\ref{fig:Pb_TNG}-a compares the $f(\NHI)$, and Figure~\ref{fig:Pb_TNG}-b and Figure~\ref{fig:Pb_TNG}-c compare the $b$ probabilities.
They are plotted in similar way to those in Figures~\ref{fig:fNHI_test2} and \ref{fig:Pb}, but for MW-like galaxies from the FEASTS sample.
Although the TNG50 predicted $f(\NHI)$ is similar to the observational one (with median offset being 0.03$\pm$0.03 when $17.8<\lgNHI<21.3$), its predicted impact parameters are too large when $\lgNHI<20$.
For a given $\lgNHI<19$, only 75\% of the predicted $\hi$ is found within the 95-percentile $b$ of the observed $\hi$, and the most extended 25\% of the predicted $\hi$ has $b$ larger by 30-50\% than that of the observed $\hi$.

\citet{Lin25b} find that TNG50 over-predicts $\hi$ radius at the $\lgNHI\sim18$ level when galaxies are massive or having low $\hi$ mass fractions within the MW-like sample (also see \citealt{Marasco25}).
Here, through the HIMF-based weighting, the $b$ probabilities have more contributions from slightly less massive galaxies than the initial MW-like subsample of FEASTS.
Yet, significant differences are still observed in the $b$ probability distributions \footnote{We also tried comparing the covering fraction of $\hi$ at the 1-kpc resolution as a function of galaxy centric radius.
The difference between TNG50 prediction and observation is less significant than the $b$ comparison.}.
Such a difference can be a mixed problem of unrealistic cool gas ($>10^4\, K$) distribution in TNG50, and imperfect $\hi$ postprocessing.

The comparisons here suggest that the conditional $b$ probability distributions at a given $\NHI$ can be very useful constraints to cosmological simulations.

\begin{figure*}
  \centering
  \includegraphics{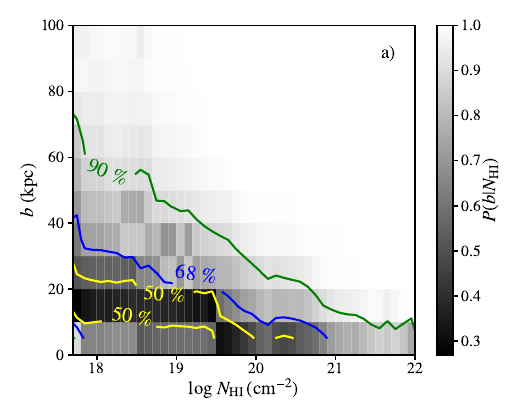}
  \includegraphics{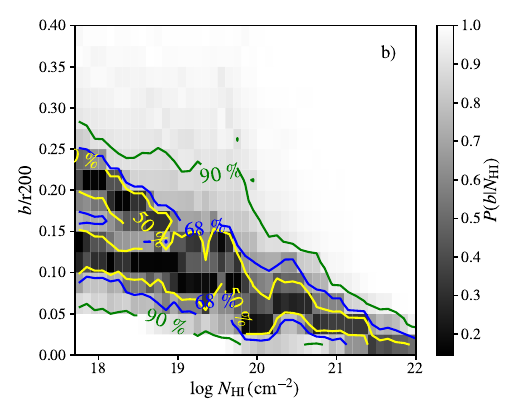}
  \includegraphics{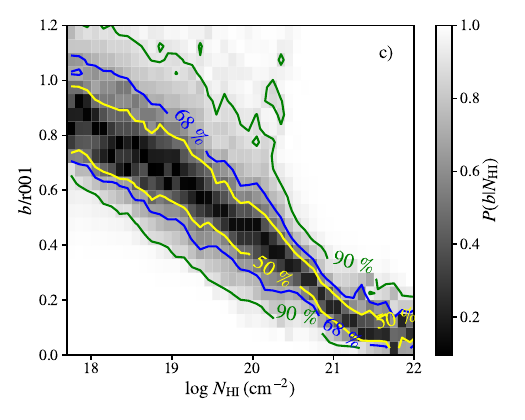}
  \includegraphics{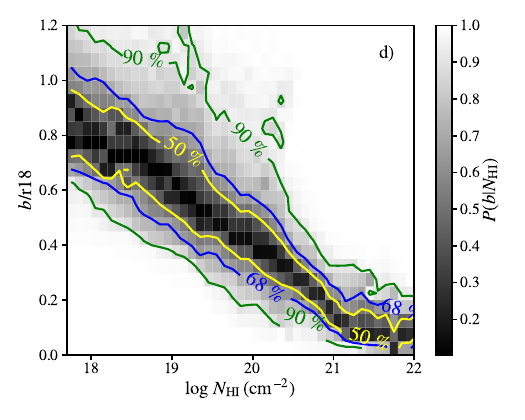}
  \caption{ Cumulative conditional probabilities of impact parameters $b$ given $\NHI$, as a function of $\NHI$ in different normalizations of $b$}.
  The probabilities are evaluated along $b$ for a given $\NHI$, and the accumulation starts from the highest probability density toward the lower ones. From panel a to c, the normalizations used are the physical unit kpc, the virial radius $r_{200}$, the $0.01\,\Msunpcsq$-$\hi$ radius $r_{001}$, and the $10^{18}\,\cmsq$-$\hi$ radius $r_{18}$, respectively. The green, blue and yellow contours show the 90, 68 and 50\% probabilities.
  \label{fig:Pb_norm}
\end{figure*}

\subsubsection{The $\NHI$ versus $b$ relations}
\label{sec:discuss_NHI_b_relation}
We investigate the probabilities of locating $\hi$ at $b$ for a given detected $\NHI$.
Figure~\ref{fig:Pb_norm} plots the relevant cumulative conditional probability distribution of $b$ given $\NHI$, as a function of $\NHI$, with $b$ normalized by different units.
The probabilities are evaluated along $b$ for a given $\NHI$, and the accumulation starts from the highest probability density toward lower values.
The radial units considered include the physical kpc,  $r_{200}$, $r_{001}$, and $r_{18}$.

The dark matter halo virial radius $r_{200}$ are estimated from the stellar mass of associated galaxies using the stellar mass-halo mass relation of \citet{Behroozi19}, with a uncertainty of $\sim$0.1 dex.
This simple estimation does not consider galaxy clustering or central versus satellite differences, and implies focus on the localized halo that is assumed to be relatively unperturbed and constrains the $\hi$ disk.
The $r_{001}$ is the radius at a projection-corrected and azimuthally averaged surface density level of 0.01 $\Msunpcsq$ ($10^{18.1}\,\cmsq$).
The $r_{18}$ is the radius at an azimuthally averaged column density of $10^{18}\,\cmsq$ without projection correction.
Because azimuthal averaging involves areas with $\lgNHI$ both below and above 18, the $r_{18}$ does not necessarily represent the average galactocentric radius to find $\lgNHI=18$ pixels.

In Figure~\ref{fig:Pb_norm}, the tightest P($b$) distribution (lowest relative scatter) is found when the normalization is with $r_{001}$ and $r_{18}$.
At $\lgNHI=17.8$,  the median $b$ in normalization of kpc, $r_{200}$, $r_{001}$, and $r_{18}$ is 30, 0.17, 0.85, and 0.82, respectively.
The 1-$\sigma$ relative scatter derived as half width between the 0.68 contours are 60, 43, 20 and 20\%, respectively.
The scatter of $b/r_{200}$ is much larger than could be produced by uncertainties of $r_{200}$.
There seems to be a bimodality in the $b$$/r_{200}$ distribution near $\lgNHI=18$, which is possibly a result of complex sample selection and will not be discussed here.

The median $b/r_{001}-\log \NHI$ relation for $\lgNHI<20.5$ is:
{ \color{green}
\begin{equation}
b/r_{001} = \lgNHI \times (-0.19\pm0.01) + (4.26\pm0.25).
\label{eq:b_N}
\end{equation}
}

Therefore, if the optical counterpart identification is secure, one can use the following procedure to estimate the $\MHI$ for each LLS associated galaxy with observed $\NHI$ and $b$.
\begin{enumerate}
\item Using the $\NHI$ value to infer the $b$$/r_{001}$ from the median relation in Figure~\ref{fig:Pb_norm}-c.
\item Combining with the observed $b$ to derive $r_{001}$.
\item Estimating $\MHI$ with $r_{001}$ based on the $r_{001}$-$\MHI$ relation \citep{Wang25}.
\end{enumerate}
The 1-$\sigma$ uncertainty of this estimate is 0.12 dex, considering both the scatter of $b$$/r_{001}$ at a given $\NHI$, and the scatter of $\MHI$ around the size-mass relation.

\subsubsection{HI covering fraction around MW-like galaxies}
The covering fraction of $\hi$ above a certain $\NHI$ within the virial radius of dark matter halos, $f_\mathrm{cov}$, has been commonly adopted in parametrizing $\hi$ distribution both in observations and in simulations \citep{Prochaska13a, Fumagalli14, Rahmati15, FaucherGiguere16, Tortora24}.
Because the $\NHI$ profiles align best with $r_{001}$ that strongly correlates with the $\MHI$ \citep{Wang25}, the $f_\mathrm{cov}$ of LLSs should strongly depend on the $\hi$ mass fraction of the central galaxy.
Therefore, comparisons of $f_\mathrm{cov}$ should only be conducted between samples with matched $\hi$ richness.

Former cosmological simulations predicted $f_\mathrm{cov}$ that was too low compared to observations at $z=3$.
This has been largely alleviated through increasing resolution and adjusting stellar feedback recipes \citep{FaucherGiguere15, FaucherGiguere16, Rahmati15, vandeVoort19}.
Modern cosmological simulations usually fail to find a correlation of $f_\mathrm{cov}$ with the halo mass, and predict very low $f_\mathrm{cov}$ at large galactic distances for MW-like halos at $z\sim0$ \citep{Tortora24}.

Our result that the $\NHI$ radial profile is mostly determined by $\MHI$ supports the lack of correlation of $f_\mathrm{cov}$ with halo mass at $z=0$.
It also hints at the possibility for a relatively low $f_\mathrm{cov}$ within the virial radius at $z\sim0$, because the general galaxy population for a halo-mass-selected MW-like sample should be relatively gas poor.
The median $\log \MHI/M_*$ of FEASTS MW-like galaxies that have $M_*\sim10^{10.5}\,M_{\odot}$ is $\sim-0.5$ (Figure~\ref{fig:scaling}), 0.8 dex higher than that of a general galaxy with the same $M_*$ \citep{Saintonge22}.
Through the $\hi$ $r_{001}$--$\MHI$ relation \citep{Wang25} and the $b$$/r_{100}$--$\NHI$ relation (equation~\ref{eq:b_N}), the median $f_\mathrm{cov}$ of the FEASTS MW-like galaxies should also be $\sim$0.8 dex higher than general MW-like galaxies.
The median $f_\mathrm{cov}$ of the FEASTS MW-like galaxies for $\lgNHI>17.8$ is $\sim$0.04 (Figure~\ref{fig:Pb_TNG}-c), for an $M_*$-based virial radius estimate of 205 kpc.
As a result, MW-like galaxies should have $f_\mathrm{cov}\sim0.006$ for $\lgNHI>17.8$, roughly in agreement with the $f_\mathrm{cov}\sim0.01$ for $\lgNHI>17.2$ at $z=0$ predicted in \citet{Tortora24} with the FIRE (Feedback In Realistic Environments, \citealt{Hopkins18}) simulation.

We emphasize that, the above deductions assume that $\hi$-poor galaxies follow similar $\NHI$ distribution as $\hi$-rich systems observed here.
We await expansion of the FEASTS sample to low $\hi$-richness systems in the near future, to confirm this result or otherwise.

\section{Summary}

We derive the $\hi$ column density distribution function $f(\NHI)$ down to a $\lgNHI\sim17.8$, from 21-cm emission line column density maps at a resolution of 1 kpc for galaxies at $z=0$.
This analysis is based on a dataset of 21-cm $\hi$ images from the FEASTS and MHONGOOSE projects, for a sample of 65 galaxies that have $\MHI$ ranging from $10^7$ to $10^{10.7}\,M_{\odot}$, and $M_*$ ranging from $10^{7.2}$ to $10^{11}\,M_{\odot}$.
We find the following when comparing to previous $z\sim0$ 21-cm emission line and quasar absorption observations:

\begin{enumerate}
\item $f(\NHI)$ can be well fitted by a Schechter function in the $\lgNHI$ range from 17.8 to 21, and is consistent with the Z05 result for the DLA regime.

\item The distribution of impact parameter $b$ to the nearest galaxy for $\lgNHI>17.8$ is tilted to significantly lower values than LLSs detected at $z<0.5$, with median $b$ being lower than 3 percentile of the LLS $b$, implying the complexity in associating absorbers to galaxies.
The TNG50 prediction overestimates the $b$ for the 25\% most distant LLS-like $\hi$ around MW-like galaxies.

\item The $\NHI$ radial distribution at $z=0$ is most strongly correlated with $r_{001}$ and $r_{18}$.
We thus propose a method to infer $\MHI$ for galaxies associated with LLSs.
\end{enumerate}

We also find the following when comparing to previous quasar absorption observations at higher $z$:

\begin{enumerate}
\item The $f(\NHI)$ at z=0 with a resolution of 1 kpc is comparable to or slightly lower that of LLS ($\sim$1 pc) at $z\geqslant3$ for $17.8<\lgNHI<19.2$, with the lower extent being larger if the cloud filling fraction is smaller.
It is systematically lower by 0.1-0.4 dex than the latter for $19.2<\lgNHI<21$.

\item The $\lgNHI>17.5$ length incidences inferred from combining the real $\hi$ images at $z=0$ with HIMFs predicted at different redshifts are systematically higher by around 1.4 times than the LLS observed ones of $0.5<z<3$.
This discrepancy indicates three possibilities of $z=0$ versus $z>0.5$ $\hi$ properties for tests in future studies: (1) the NUM predicted HIMFs under-estimates the galaxy densities by 1.4 times at $z>0.5$, (2) $z=0$ LLS cross-sections are 1.4 times higher than at higher redshift, (3) the average covering fraction of LLS clouds in 1-kpc areas that have LLS-like column densities is 70\%.
\end{enumerate}

\section*{acknowledgments}
We gratefully thank X. Z. Chen for useful discusisons.
JW thanks support of the research grants from Ministry of Science and Technology of the People's Republic of China (NO. 2022YFA1602902),  the National Natural Science Foundation of China (NO. 12073002), and the science research grants from the China Manned Space Project (NO. CMS-CSST-2021-B02).
This work has received funding from the European Research Council (ERC) under the European Union's Horizon 2020 research and innovation programme (grant agreement No. 882793 ``MeerGas'').
LCH was supported by the National Natural Science Foundation of China (11721303, 11991052, 12011540375, 12233001), the National Key R\&D Program of China (2022YFF0503401).
PK is partially supported by the BMBF project 05A23PC1 for D-MeerKAT.
Parts of this research were supported by High-performance Computing Platform of Peking University.
DL thanks grants of NSFC (12588202).
KN is supported by the MEXT/JSPS KAKENHI Grant Numbers JP22K21349, 24H00002, 24H00241 and 25K01032.

This work made use of the data from FAST (Five-hundred-meter Aperture Spherical radio Telescope, \url{https://cstr.cn/31116.02.FAST}). FAST is a Chinese national mega-science facility, operated by National Astronomical Observatories, Chinese Academy of Sciences.

\facilities{FAST: 500 m, VLA, WSRT, MeerKAT }
\software{Astropy \citep{astropy:2013, astropy:2018, astropy:2022},
numpy \citep[v1.21.4]{vanderWalt11}, photutils \citep[v1.2.0]{Bradley19}, Python \citep[v3.9.13]{Perez07}, scipy \citep[1.8.0]{Virtanen20} }

\appendix

\section{Table of the interferometric data for the FEASTS sample}
\begin{figure*}
  \centering
  \includegraphics{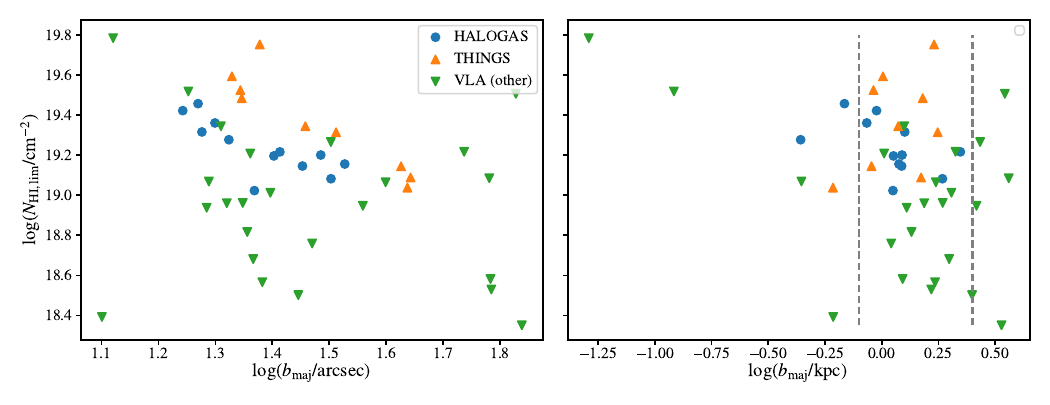}
  \caption{ Depth and resolution of interferometry data for galaxies in the initial FEASTS+interferometric sample. The circles, upward triangles, and downward triangles are for data from the HALOGAS sample, the THINGS sample, and other VLA observed data. The left panel is a function of angular resolution; the right panel shows physical resolution. The two dashed vertical lines in the right panel enclose the range selected by this study, 0.8 to 2.4 kpc. Samples to the left of this range are downgraded to the resolution of 0.8 kpc, and samples to the right of this range are excluded from the final FEASTS sample.
 }
  \label{fig:intprop}
\end{figure*}

We show in Figure~\ref{fig:intprop} column density sensitivity as a function of angular and physical resolution for the initial FEASTS+interferometric sample of 41 galaxies.
We present in Table~\ref{tab:intdata} the properties of the interferometric data collected from past surveys or archives of interferometric telescopes for the finally analyzed FEASTS sample of 35 galaxies.

\begin{table*}
  \centering
          \caption{The interferometric data for the FEASTS sample (40 galaxies)}
          \begin{tabular}{l l l l l l}
            \hline

            Name &  Survey or archive &  proj  &  array  & $b_\mathrm{maj}$ & $\log\,N_\mathrm{HI,lim}$  \\
                  &    &   &    & ($"$) & ($\mathrm{cm}^{-2}$)  \\
            (1)     & (2)     & (3)  & (4)  & (5) & (6)   \\
            \hline

IC1613 & {\scriptsize LITTLE THINGS} & - & - & 13.2 & 19.78 \\
 NGC1055 & VLA & AG512 & C ($6.4\,h$)  & 22.7 & 18.82 \\
NGC2541 & HALOGAS & - & - & 17.5 & 19.42 \\
NGC2841 & THINGS & - & - & 10.4 & 20.08 \\
NGC2903 & THINGS & - & - & 16.5 & 19.73 \\
NGC3198 & HALOGAS & - & - & 18.9 & 19.32 \\
NGC3344 & VLA & 20A-125(B/C),21A-408(D)  & B ($27.8\,h$), C ($9.6\,h$), D ($2.5\,h$)  & 12.6 & 18.39 \\
NGC3486 & VLA & 13B-363 (B), 13A-107 (C) &  B ($2.7 \,h$)  C ($6.6 \,h$) & 24.1 & 18.57 \\
NGC3521 & THINGS & - & - & 15.9 & 19.88 \\
NGC4214 & THINGS & - & - & 16.0 & 19.68 \\
NGC4244 & HALOGAS & - & - & 21.1 & 19.28 \\
NGC4254 & VIVA & - & - & 24.9 & 19.01 \\
NGC4258 & HALOGAS & - & - & 18.6 & 19.46 \\
NGC4395 & VLA & AW619 & B ($19.7\,h$),  C ($8.0\,h$) & 19.4 & 19.07 \\
NGC4414 & HALOGAS & - & - & 25.9 & 19.22 \\
 NGC4449 & VLA & AH540  & D ($2.1\,h$) & 60.7 & 18.58 \\
 NGC4490 & VLA & AA181 &  C ($5.9\,h$), D ($4.7\,h$) & 23.0 & 19.21 \\
 NGC4527 & VLA & AM788 & C ($4.1\,h$) & 23.2 & 18.68 \\
NGC4532 & VIVA & - & - & 19.3 & 18.94 \\
 NGC4536 & VLA & AN119 & C ($6.8\,h$) & 20.9 & 18.96 \\
NGC4559 & HALOGAS & - & - & 28.4 & 19.15 \\
NGC4565 & HALOGAS & - & - & 31.8 & 19.08 \\
 NGC4618 & VLA &  AW618 & B ($14.3\,h$),  C ($15.5\,h$)  & 29.5 & 18.76 \\
NGC4631 & HALOGAS & - & - & 33.7 & 19.16 \\
NGC4656 & VLA & AR215  & D ($4.8\,h$)  & 54.6 & 19.22 \\
NGC4826 & THINGS & - & - & 11.6 & 20.13 \\
NGC5033 & VLA &  AP270 & C ($9.4 \,h$) D ($4.6 \,h$) & 27.9 & 18.50 \\
NGC5055 & HALOGAS & - & - & 19.9 & 19.36 \\
NGC5194 & THINGS & - & - & 11.0 & 20.09 \\
 NGC5248 & VLA & AS787 & C ($2.9\,h$) & 20.4 & 19.34 \\
NGC5457 & THINGS & - & - & 13.3 & 20.02 \\
 NGC5474 & VLA & AH124 & D ($4.8\,h$) & 60.9 & 18.53 \\
 NGC5907 & VLA & 12A-405 (B), AS608 (C), AR302 (D)  &  B ($21.2 \, h$) C ($4.7 \,h$) D ($0.6 \,h$) 22.3 & 18.96 \\
NGC628 & THINGS & - & - & 11.0 & 20.16 \\
 NGC660 & VLA &  AG230 & C ($7.2\,h$), D($6.6\,h$)  & 39.7 & 19.06 \\
NGC672 & HALOGAS & - & - & 30.6 & 19.20 \\
NGC7331 & THINGS & - & - & 7.5 & 20.50 \\
NGC891 & HALOGAS & - & - & 23.4 & 19.02 \\
NGC925 & HALOGAS & - & - & 25.3 & 19.20 \\
SexB & {\scriptsize LITTLE THINGS} & - & - & 17.9 & 19.52 \\

            \hline
          \end{tabular}

            \raggedright
              Column~(1): Galaxy name.
              Column~(2): Survey name or the VLA archive, as orgin of the interferometric data.
              Column~(3): The project id of the data if the origin is the VLA archive.
              Column~(4): The array configruations of data used, if the origin is the VLA archive. The on-source time is denoted in bracket after each configuration.
              Column~(5): The major axis of the synthesized beam.
              Column~(6): The column density limit (3-$\sigma$), quantified as the level of the $\lgNHI$ contour where the median $\SN=3$.
      \label{tab:intdata}
  \end{table*}

\section{Verifying the FAST beam unification and clean procedure with mock tests}
\label{app:mock}
The multi-beam receivers on FAST and many other single-dish radio telescopes have significantly accelerated imaging surveys, through scanning the skys in drifting or on-the-fly mode.
But when targeting low-surface-density structures, the images commonly have the problem of each beam of the multi-beam responds to the sky with different scattering patterns.
They further share the problem of single-beam receivers that the beam response have side lobes and thus the shape deviates from a perfect Gaussian function, because of the blocking and diffraction of light.
These problems hinder the scientific interpretation of low-surface-density structures detected in images.
We have thus design procedures to unify and clean the beam for the diffuse $\hi$ images in section~\ref{sec:combine_feasts_traditional}.
The diffuse $\hi$ is the excess $\hi$ detected in a FAST image with respect to the corresponding interferoemtric image, and adding the diffuse $\hi$ image to the interferometric image produces the full $\hi$ image.

In the following, we conduct mock tests to verify the robustness of the beam unification and clean procedures in the final construction of full $\hi$ images.

\begin{figure*}
  \centering
    \includegraphics[width=\textwidth*4/9]{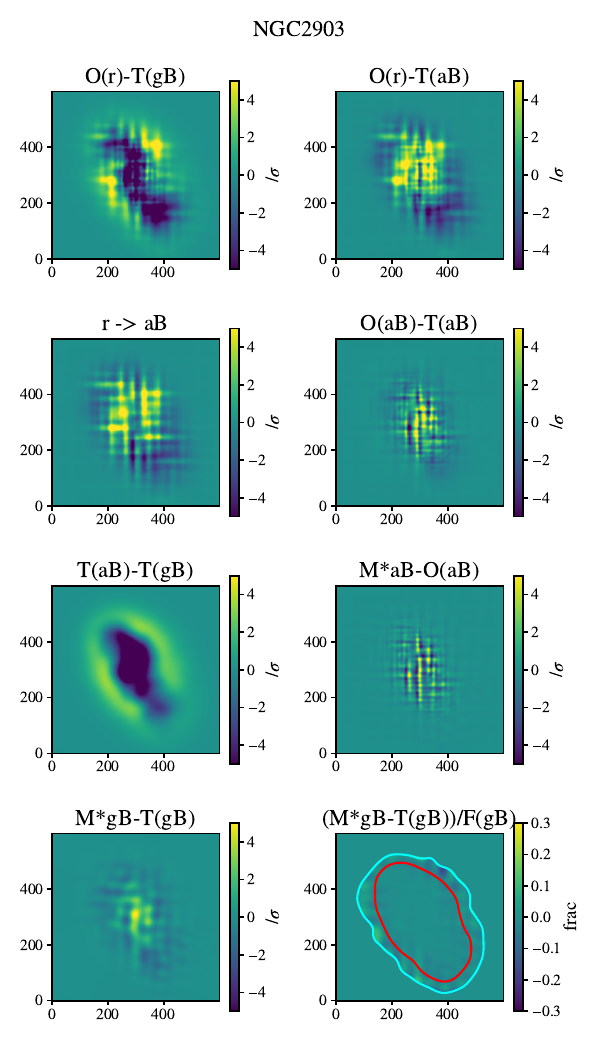}
    \includegraphics[width=\textwidth*4/9]{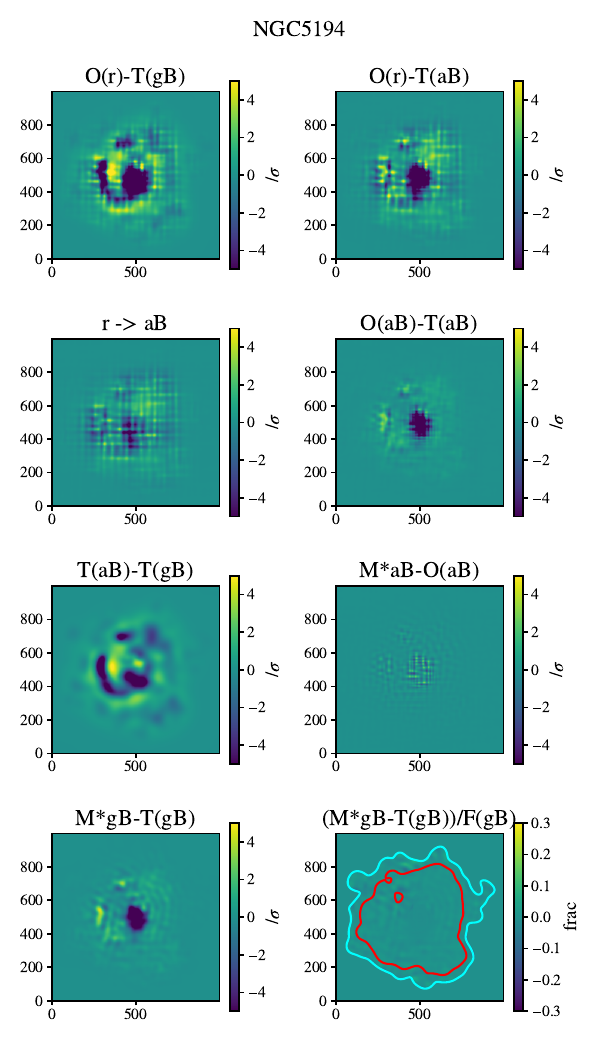}
  \caption{Example mock test galaxies, for which the input interferometric data are from the VLA. The left and right two columns are for two mock galaxies respectively. Please refer to Appendix~\ref{app:mock} for detailed explanation of the image titles.
 }
  \label{fig:QA1}
\end{figure*}

We simulate the whole full-$\hi$ image constructing process, with a focus on correctly producing the diffuse $\hi$ image.
The full-$\hi$ image (dense$+$diffuse $\hi$) and the interferometric convolved model (dense $\hi$) of each galaxy in the FEASTS sample are converted to have unit of Jy$/$arcsec, and used as the input images.
They represent skies seen in the single-dish and interferometric mock observations, respectively.
The mock observation, data processing, and final verification are conducted with the procedure described below.
We use mock images built with the data of NGC 2903 as an example (left column of Figure~\ref{fig:QA1}).

The input full-$\hi$ image enters the FAST observing simulator, with noise added, and produces the mock observed FAST image.
It is also directly convolved with the average beam (uB), and a Gaussian beam that has FWHM equal to 3$'$.24 (gB), producing the reference FAST images.
The input interferometric convolved model is convolved with the same synthesized interferometric beam, producing the mock observational interferometric image.
By doing so, we miss the densest $\hi$ in the original data, but this $\hi$ does not affect processing of the diffuse $\hi$.
The three types of FAST images described above subtracting the mock observational interferometric image produce three observed diffuse $\hi$ images.
The first one mimics the {\it observational} data with a {\it varied} and non-Gaussian beam (O(vB)), which is to be corrected.
The remaining two images serve as the reference true images, one has a {\it uniform} but non-Gaussian beam (T(uB)), and the other has a uniform and {\it Gaussian beam} (T(gB)).
They are used to check the quality of beam unification and cleaning, respectively.
In Figure~\ref{fig:QA1}, we can see that O(vB) deviate significantly from both T(uB) and T(gB).

We apply to O(vB) the beam unification procedure described in section~\ref{sec:combine_feasts_traditional}, and obtain the beam unified image (O(uB)), as well as the correction image (C(vB$\rightarrow$uB)).
Figure~\ref{fig:QA1} shows that, C(vB$\rightarrow$uB) has similarly horizontal and vertical stripes as O(vB)-T(uB), but these artifical patterns mostly disappear in O(uB)-T(uB), indicating the success of beam unification.
But T(uB)-T(gB) shows significant none-zero patterns, suggesting the neccessity of applying the beam clean procedure.

We apply to O(uB) the beam clean procedure described in section~\ref{sec:combine_feasts_traditional}, and obtain the clean model (M).
The clean residual, M*uB-O(uB), is relatively flat, indicating the success of deconvolution.
The restored clean image M*gB does not perfectly recover the T(gB), if the absolute residual is compared to the FAST image rms, reflecting challenges in deconvolution.
However, the fraction residual with respect to fluxe in the reference full-$\hi$ image (F(gB)=T(gB)+mock observational interferometric image) is relatively small, mostly within 10\% (or 0.04 dex).

\begin{figure*}
  \centering
    \includegraphics[width=\textwidth*4/9]{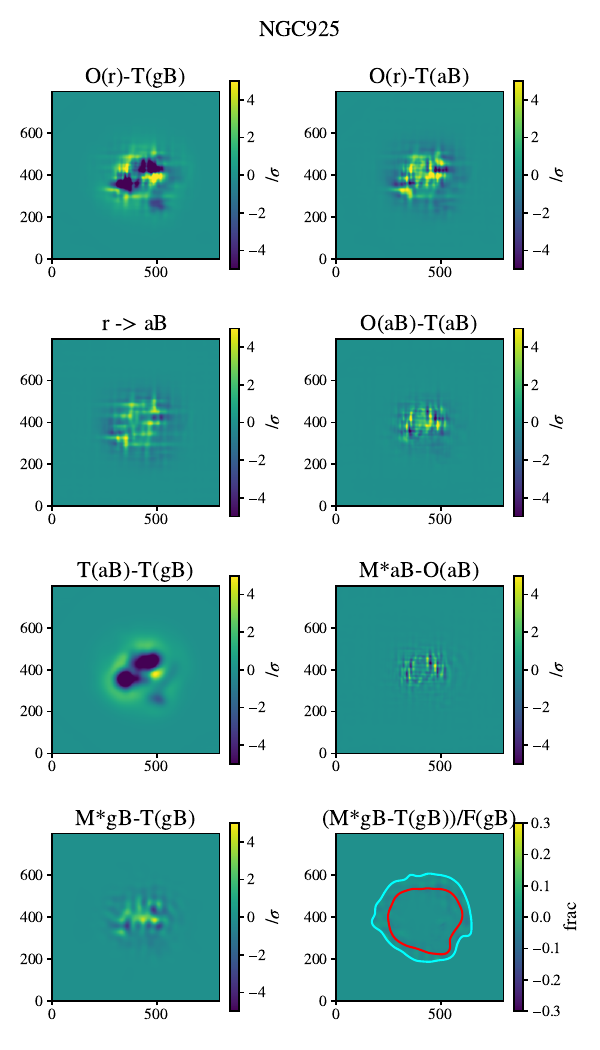}
    \includegraphics[width=\textwidth*4/9]{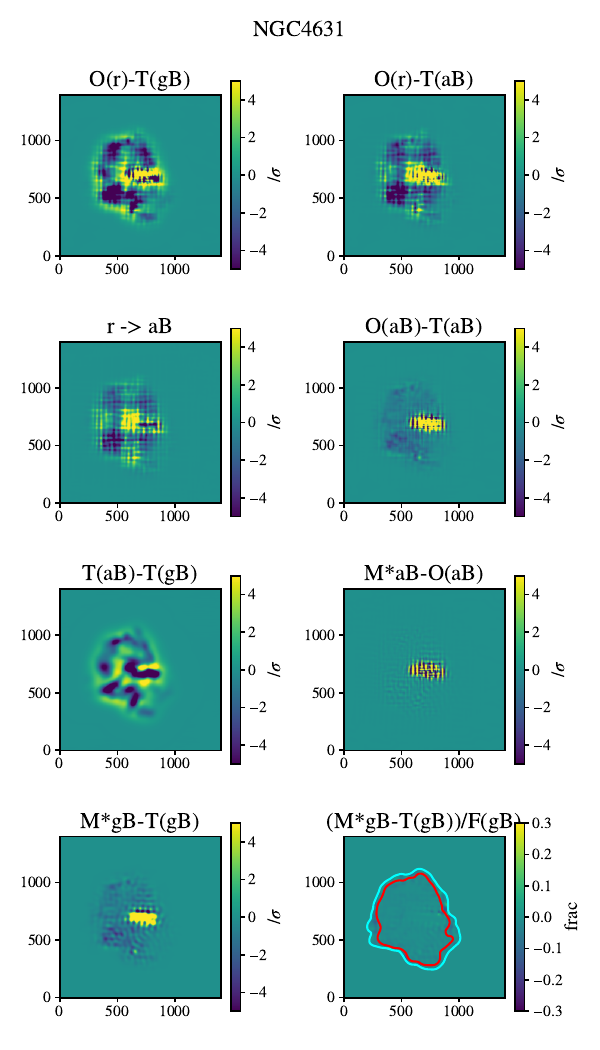}
  \caption{ Example mock test galaxies. Similar to Figure~\ref{fig:QA1}, but with the input interferometric data are from the WSRT (HALOGAS).
 }
  \label{fig:QA2}
\end{figure*}

We repeat the procedure above for all galaxies in the FEASTS sample, and reach similar conclusions.
Three more examples are presented in the right column of Figure~\ref{fig:QA1}, and in Figure~\ref{fig:QA2}.
The four example galaxies represent relatively isolated and interacting galaxies with interferometric data from the VLA (Figure~\ref{fig:QA1}) and WSRT (where HALOGAS was conducted, Figure~\ref{fig:QA2}), respectively.
The NGC 4631 example shows the worst type of clean residual encountered in the whole sample, possibly because it is a strongly interacting edge-on galaxy that has a complex morphology.
Yet the fractional residuals with respect to full-$\hi$ are acceptable.

The diffuse $\hi$ clean residuals discussed above are also the differences between recovered and reference full-$\hi$ images, because the mock interferometric images remain the same.
Because the $f(\NHI)$ counts areas in $\lgNHI$ bins of 0.1-dex wide, the image recovery with typical systematic uncertainty less than 10\% is satisfactory.
Such a conclusion is further supported by the high level of consistency between the $f(\NHI)$ derived with reference full-$\hi$ images (true) and that with recovered full-$\hi$ images (measure) in Figure~\ref{fig:fNHI_test1}.

\section{The $f(\NHI)$ data points and Schechter function fitting}
\label{app:fNHI}
Table~\ref{tab:fNHI} lists the data points and error bars for the $f(\NHI)$ derived in this study.

\begin{figure*}
  \centering
  \includegraphics[width=\textwidth]{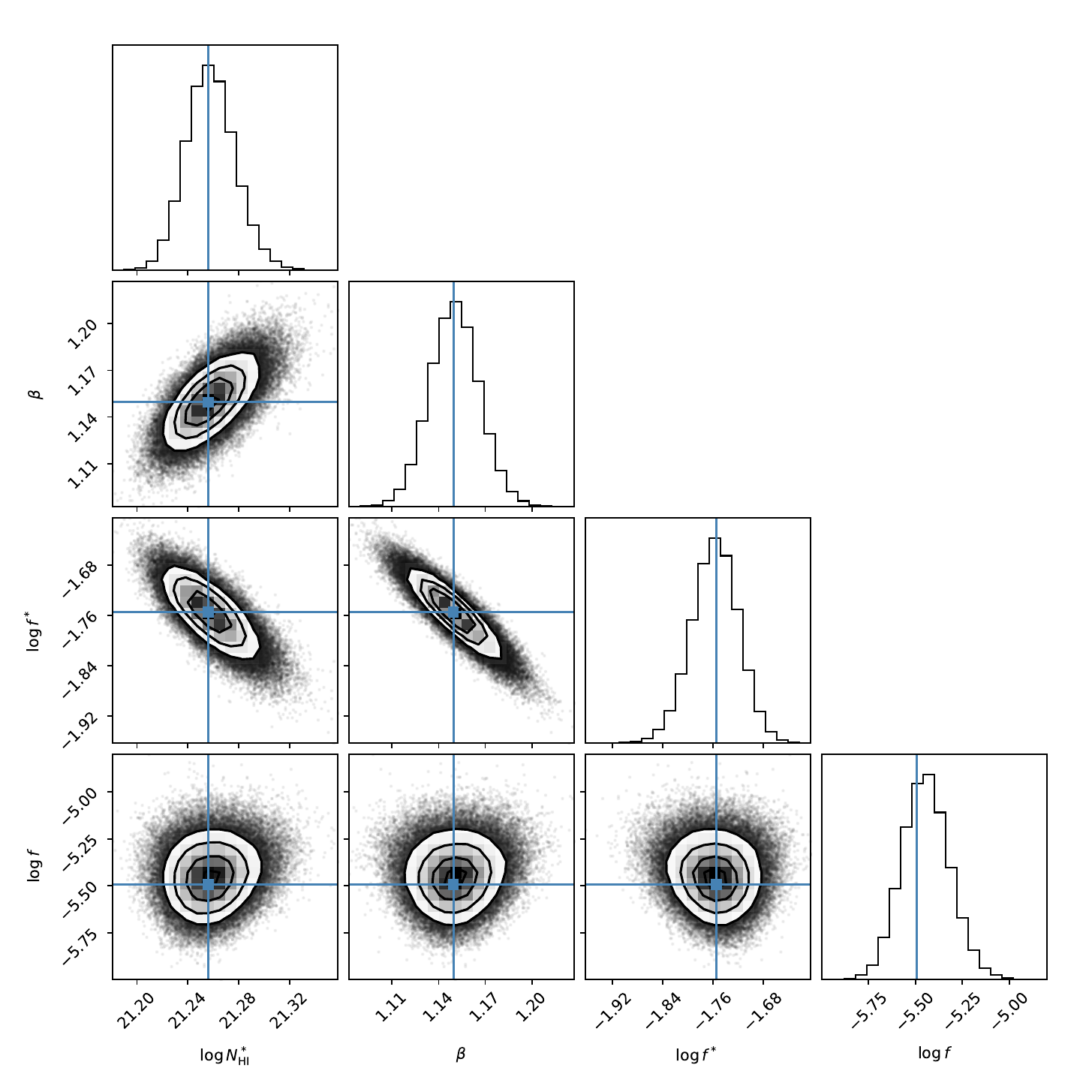}
  \caption{Schechter function fitting for the $f(\NHI)$ measurements. The parameters $\lg \NHI^*$, $\beta$ and $\log f^*$ together form the Schechter function displayed in Eqation~\ref{eq:fNHI}. The $f$ of $\log f$ describes the fractional level of variance underestimation. It plots all the one and two dimensional projections of the posterior probability distributions of parameters, with prior distributions described in the main text (Section~\ref{sec:result}). The best-fit values (from scipy.optimize) are marked as vertical and horizontal lines.}
  \label{fig:mcmc}
\end{figure*}

Figure~\ref{fig:mcmc} shows the corner figure for the Schechter function fitting generated by the emcee procedure \citep{ForemanMackey13}.
It displays probability distributions in the parameter space, and the weak degeneracy between parameters.
The final parameter $f$ is the fractional systematic uncertainties of the model in describing the data, which is very small.

\begin{table}
  \centering
          \caption{The HI column density distribution function}
          \begin{tabular}{c c c }
            \hline
            $\lgNHI$ &  $\log\,f(\NHI)$  &  $\sigma(\log\,f(\NHI))$  \\
                    &    &     (dex)   \\
            (1)     & (2)     & (3)   \\
            \hline
          17.75 & -19.20 & 0.06 \\
          17.85 & -19.17 & 0.04 \\
          17.95 & -19.26 & 0.04 \\
          18.05 & -19.36 & 0.04 \\
          18.15 & -19.47 & 0.04 \\
          18.25 & -19.57 & 0.04 \\
          18.35 & -19.68 & 0.04 \\
          18.45 & -19.78 & 0.05 \\
          18.55 & -19.88 & 0.04 \\
          18.65 & -19.99 & 0.04 \\
          18.75 & -20.10 & 0.04 \\
          18.85 & -20.20 & 0.04 \\
          18.95 & -20.29 & 0.04 \\
          19.05 & -20.39 & 0.04 \\
          19.15 & -20.48 & 0.04 \\
          19.25 & -20.62 & 0.05 \\
          19.35 & -20.77 & 0.04 \\
          19.45 & -20.93 & 0.05 \\
          19.55 & -21.05 & 0.04 \\
          19.65 & -21.15 & 0.04 \\
          19.75 & -21.26 & 0.04 \\
          19.85 & -21.37 & 0.04 \\
          19.95 & -21.48 & 0.04 \\
          20.05 & -21.56 & 0.04 \\
          20.15 & -21.65 & 0.04 \\
          20.25 & -21.86 & 0.04 \\
          20.35 & -22.03 & 0.04 \\
          20.45 & -22.19 & 0.04 \\
          20.55 & -22.35 & 0.05 \\
          20.65 & -22.45 & 0.06 \\
          20.75 & -22.57 & 0.05 \\
          20.85 & -22.67 & 0.05 \\
          20.95 & -22.80 & 0.05 \\
          21.05 & -22.98 & 0.05 \\
          21.15 & -23.25 & 0.04 \\
          21.25 & -23.53 & 0.05 \\
          21.35 & -23.79 & 0.08 \\
          21.45 & -24.11 & 0.10 \\
          21.55 & -24.47 & 0.12 \\
          21.65 & -24.82 & 0.16 \\
          21.75 & -24.98 & 0.18 \\
          21.85 & -25.23 & 0.20 \\
          21.95 & -25.80 & 0.28 \\
          22.05 & -26.30 & 0.29 \\
          22.15 & -28.42 & 1.32 \\
            \hline
          \end{tabular}
            \raggedright
              Column~(1): the center of $\hi$ column density bins.
              Column~(2): the $\hi$ column density distribution function.
              Column~(3): the error of the $\hi$ column density distribution function.
      \label{tab:fNHI}
  \end{table}

\bibliography{fNHI}{}
\bibliographystyle{apj}

\end{document}